%% file: main.tex
\PassOptionsToPackage{dvipsnames}{xcolor}
\documentclass[a4paper,UKenglish,cleveref, autoref,thm-restate]{lipics-v2021}

\title{Structural Parameterization of Locating-Dominating Set and Test Cover} 

\titlerunning{Structural Parameterization of Locating-Dominating Set and Test Cover} 

\author{Dipayan Chakraborty}{Université Clermont Auvergne, CNRS, Mines Saint-Étienne, Clermont Auvergne INP, LIMOS, 63000 Clermont-Ferrand, France\and Department of Mathematics and Applied Mathematics, University of Johannesburg, Auckland Park, 2006, South Africa \and \url{https://dipayan5186.github.io/Website/}}{dipayan.chakraborty@uca.fr}{https://orcid.org/0000-0001-7169-7288}{Public grant overseen by the French National Research Agency as part of the “Investissements d’Avenir” through theIMobS3 Laboratory of Excellence (ANR-10-LABX-0016), the International Research Center ``Innovation Transportation and Production Systems'' of the I-SITE CAP 20-25 and by the ANR project GRALMECO (ANR-21-CE48-0004).}

\author{Florent Foucaud}{Université Clermont Auvergne, CNRS, Mines Saint-Étienne, Clermont Auvergne INP, LIMOS, 63000 Clermont-Ferrand, France\and \url{https://perso.limos.fr/ffoucaud}}{florent.foucaud@uca.fr}{https://orcid.org/0000-0001-8198-693X}{ANR project GRALMECO (ANR-21-CE48-0004) and the French government IDEX-ISITE initiative 16-IDEX-0001 (CAP 20-25).}

\author{Diptapriyo Majumdar}{Indraprastha Institute of Information Technology Delhi, Delhi, India\and \url{https://diptapriyomajumdar.wixsite.com/toto}}{diptapriyo@iiitd.ac.in}{https://orcid.org/0000-0003-2677-4648}{Supported by the Science and Engineering Research Board (SERB) Grant SRG/2023/001592}

\author{Prafullkumar Tale}{Indian Institute of Science Education and Research Bhopal, Bhopal, India \and \url{https://pptale.github.io/}}{prafullkumar@iiserb.ac.in}{https://orcid.org/0000-0001-9753-0523}{Supported by INSPIRE Faculty Fellowship by DST, India and
Startup Research Grant by IISER-Bhopal.}

\authorrunning{Chakraborty, Foucaud, Majumdar, and Tale} 

\Copyright{Chakraborty, Foucaud, Majumdar, Tale} 

\ccsdesc[500]{Theory of computation~Fixed parameter tractability}

\keywords{Identification Problems, Locating-Dominating Set, Test Cover, Parameterized Algorithms, Structural Parameterizations, Kernelization} 

\category{} 

\relatedversion{} 

\EventEditors{John Q. Open and Joan R. Access}
\EventNoEds{2}
\EventLongTitle{42nd Conference on Very Important Topics (CVIT 2016)}
\EventShortTitle{CVIT 2016}
\EventAcronym{CVIT}
\EventYear{2016}
\EventDate{December 24--27, 2016}
\EventLocation{Little Whinging, United Kingdom}
\EventLogo{}
\SeriesVolume{42}
\ArticleNo{23}

\input{preamble}

\begin{document}

\maketitle

\begin{abstract}
We investigate structural parameterizations for two identification problems
in graphs and set systems:
\textsc{Locating-Dominating Set} and \textsc{Test Cover}.
In the first problem, an input is a graph $G$ on $n$ vertices and an
integer $k$, and one asks whether there is a subset $S$ of $k$
vertices such that any two distinct vertices not in $S$ are dominated
by distinct subsets of $S$. In the second problem, an input is a set
of items $U$, a collection of subsets $\calF$ of $U$
called \emph{tests}, and an integer $k$, and one asks whether there is
a solution set $S$ of at most $k$ tests such that each pair of items
belongs to a distinct subset of tests of $S$. 
These two problems are ``identification'' analogues of the
\textsc{Dominating Set} and \textsc{Set Cover} problems, respectively.

Chakraborty et al. [ISAAC 2024] proved,
among other things, that both these problems admit a conditional
double-exponential lower bound and a matching algorithm when parameterized by the treewidth of the input graph.
We continue this line of investigation and consider parameters
larger than the treewidth, like the vertex cover number and feedback edge set number.
We design a nontrivial dynamic programming scheme to
solve \textsc{Test Cover} in ``slightly super-exponential'' time
$2^{\calO(|U|\log |U|)}(|U|+|\calF|)^{\calO(1)}$ in the number $|U|$
of items, and also \textsc{Locating-Dominating Set} in time
$2^{\calO(\vc\log\vc)} \cdot n^{\calO(1)}$, where $\vc$ is the vertex cover number and $n$ the order of the graph.
We thus show that the lower bound results with respect to
treewidth from Chakraborty et al. [ISAAC 2024] 
cannot be extended to the vertex cover number.
We also show that when parameterized by the feedback edge set number,  
\textsc{Locating Dominating Set} admits a linear kernel,
thereby answering an open question from [Cappelle et al., LAGOS 2021].
Finally, we show that neither \textsc{Locating-Dominating Set} nor \textsc{Test Cover} is likely to admit a compression algorithm returning an input with a subquadratic number of bits,
unless $\NP \subseteq \coNP/poly$.
\end{abstract}

\input{introduction}
\input{prelims}
\input{vc-fpt}

\input{other-structural-para}
\input{fes-kernel}
\input{incompressibility}
\input{conclusion}

\bibliography{references}

\end{document}

%% file: preamble.tex
\usepackage[linesnumbered,algoruled,boxed,lined]{algorithm2e}
\usepackage{complexity} 
 
\usepackage{tikz}
\usetikzlibrary{positioning,backgrounds,patterns,calc}

\usepackage{enumitem}

\usepackage{thmtools}
\usepackage{thm-restate}

\usetikzlibrary{patterns}

\usepackage{subcaption}
\usepackage{enumitem}
 
 \usepackage{graphicx}
 \usepackage{amsmath}
 \usepackage{amsthm}

\usepackage[dvipsnames]{xcolor}

\newcommand{\pt}[1]{\textcolor{purple}{#1 - PT}}

\newcommand{\LD}{\textsc{Locating-Dominating Set}\xspace}
\newcommand{\TCPB}{\textsc{Test Cover}\xspace}

\newcommand{\bigo}{\mathcal{O}}

\newcommand{\bitrep}{\texttt{bit-rep}}
\newcommand{\bit}{\texttt{bit}}

\newcommand{\ETH}{{\textsf{ETH}}}

\newcommand{\tw}{\textsf{tw}}

\newcommand{\fes}{\textsf{fes}}
\newcommand{\vc}{\textsf{vc}}

\newcommand{\tc}{\textsf{tc}}
\newcommand{\dclique}{\textsf{dc}}

\newcommand{\calA}{\mathcal{A}}
\newcommand{\calB}{\mathcal{B}}

\newcommand{\calF}{\mathcal{F}}

\newcommand{\calO}{\mathcal{O}}
\newcommand{\calP}{\mathcal{P}}

\newcommand{\calX}{\mathcal{X}}

\newcommand{\yes}{\textsc{Yes}}
\newcommand{\no}{\textsc{No}}

\newcommand{\x}{\mathcal{X}}

\newcommand{\q}{\mathcal{Q}}

\newcommand{\lc}{\texttt{opt}}
\newtheorem{reduction rule}[theorem]{Reduction Rule}
\nolinenumbers

\newcommand{\defproblem}[3]{
  \vspace{1mm}
\noindent\fbox{
  \begin{minipage}{0.96\textwidth}
  \begin{tabular*}{\textwidth}{@{\extracolsep{\fill}}lr} #1 \\ \end{tabular*}
  {\bf{Input:}} #2  \\
  {\bf{Question:}} #3
  \end{minipage}
  }
  \vspace{1mm}
}


%% file: introduction.tex
\section{Introduction}

We study two discrete identification problems, \LD\ and \TCPB, that are extensively studied since the 1970s. This
type of problems, in which one aims at distinguishing all the elements of a discrete structure by a solution set, is popular in both
combinatorics~\cite{bondy1972induced,renyi1961} and
algorithms~\cite{BDK05,BHHHLRS03}, and has many natural applications
e.g. to network monitoring~\cite{Rao93}, medical diagnosis~\cite{MS85}
or machine learning~\cite{CN98}. In a recent paper, Chakraborty et
al.~\cite{DBLP:journals/corr/abs-2402-08346} revisited the
parameterized complexity of \LD\ and \TCPB\ when parameterized by {the
  solution size or the treewidth ($\tw$) of the input graph.  In the
  case of \TCPB, the authors considered the treewidth of the natural
  auxiliary bipartite graph (which we specify later).  They proved
  that \LD\ and \TCPB\ admit an algorithm running in time
  $2^{2^{\calO(\tw)}} \cdot {\rm poly}(n)$, where $n$ is the number of
  elements/vertices in the system/graph.  More interestingly, they proved that these
  algorithms are tight, i.e., neither problem admits an algorithm
  running in time $2^{2^{o(\tw)}} \cdot {\rm poly}(n)$, unless the
  \ETH\ fails.  These results add \LD\ and \TCPB\ to the small list of
  \NP-complete problems, recently initiated by Foucaud et
  al. \cite{DBLP:conf/icalp/FoucaudGK0IST24}, that admit a conditional
  double-exponential lower bound when parameterized by the treewidth.
  We continue this line of research and study the
  parameterized complexity of these problems for parameters
  that are larger than the treewidth.  Before moving forward, we formally
  define the two studied problems.

\defproblem{\LD}{A graph $G$ on $n$ vertices and an integer $k$.}{Does 
there exist a locating-dominating set of size $k$ in $G$, that is,
a dominating set $S$ of $G$ such that for any two different 
vertices $u, v \in V(G) \setminus S$,
their neighbourhoods in $S$ are different, 
i.e., $N(u) \cap S \neq N(v) \cap S$?} 

\defproblem{\TCPB}{A set of items $U$, a collection of subsets of $U$ called \emph{tests} and denoted by $\calF$, and an integer $k$.}{Does 
there exist a collection of at most $k$ tests
such that for each pair of items, there is a test 
that contains exactly one of the two items?}

\noindent As \TCPB\ is defined over set systems, 
to consider structural graph parameters, we define the \emph{auxiliary graph}
of a set system $(U,\calF)$ in the natural way:
The bipartite graph $G$ on $n$ vertices with bipartition $\langle R, B \rangle$
of $V(G)$ such that sets $R$ and $B$ contain a vertex
for every set in $\calF$ and for every item in $U$, respectively,
and where $r \in R$ and $b \in B$ are adjacent 
if and only if the set corresponding to $r$ contains 
the element corresponding to $b$.

We refer readers to~\cite{DBLP:journals/corr/abs-2402-08346} for a comprehensive overview about 
the motivations, applicability and known literature regarding 
the two problems,
as we only mention the most relevant results here.
\LD~\cite{colbourn1987locating} and \TCPB~\cite[SP6]{GJ79} are both \NP-complete.
They are also trivially \FPT\ for the
parameter solution size (see~\cite{DBLP:journals/corr/abs-2011-14849,CGS21,CGJSY12} or the discussion before 
Theorem~$2$ and Theorem~$3$ in \cite{DBLP:journals/corr/abs-2402-08346}).
Hence, it becomes interesting to study these problems under
more refined angles, such as kernelization, fine-grained complexity,
and alternative parameterizations. 
Naturally, these lines of research 
have been pursued in the literature. 
\TCPB was studied within the framework of 
``above/below guarantee'' parameterizations 
in~\cite{BFRS16,CGJMY16,CGJSY12,GMY13} and kernelization 
in~\cite{BFRS16,CGJMY16,GMY13}. 
These results have shown an intriguing 
behaviour for \TCPB, with some nontrivial techniques being developed
to solve the problem~\cite{BFRS16,CGJSY12}. 
In the case of \LD, fine-grained complexity results regarding the number of vertices 
and solution size
were obtained in~\cite{BIT20,DBLP:journals/corr/abs-2011-14849,CGS21}, and structural parameterizations of the input graph 
have been studied in~\cite{DBLP:journals/corr/abs-2011-14849,CGS21}. 
In particular, it was shown there that the problem admits a linear
kernel for the parameter max leaf number, however (under standard
complexity assumptions) no polynomial kernel exists for the solution
size, combined with either the vertex cover number or the distance to
clique. They also provide a double-exponential kernel for the
parameter distance to cluster.

\subparagraph{Our Contributions.}
We extend the above results by systematically studying the
fine-grained complexity and kernelization of \LD and \TCPB for
the parameters number of vertices/elements/tests as well as
structural parameters.
Our results contribute to the complexity landscape for structural parameterizations of 
\LD started in~\cite{DBLP:journals/corr/abs-2011-14849,CGS21}. See Figure~\ref{fig:result-overview} for an overview of the known results
for (structural) parameterizations of \LD, using a 
Hasse diagram of standard graph parameters.

Our first result is an FPT algorithm for \LD parameterized by 
the vertex cover number (${\vc}$) of the input graph, and 
for \TCPB when parameterized by the number $|U|$ of items.

\begin{restatable}{theorem}{locdomvertexcoveralgo}
\label{thm:LD-struct-vc}
\LD admits an algorithm running in time 
$2^{\calO(\vc\log\vc)} \cdot n^{\calO(1)}$, 
where \vc\ is the vertex cover number of the input graph.
Also, \TCPB admits an algorithm running in time 
$2^{\calO(|U|\log|U|)} \cdot (|U| + |\calF|)^{\calO(1)}$.
\end{restatable}

The above result shows that unlike treewidth, 
for which we cannot expect a running time of the form $2^{2^{o({\tw})}}$~\cite{DBLP:journals/corr/abs-2402-08346}, 
if we choose ${\vc}$, a parameter larger than treewidth,
then we can significantly improve the running time.
Moreover, we show that the above result for \LD\ also extends 
to the parameters distance to clique and twin-cover number.
For \TCPB, the result is reminiscent of the known $\calO(2^n mn)$ 
dynamic programming scheme for \textsc{Set Cover} 
on a universe of size $n$ and $m$ sets 
(see~\cite[Theorem 3.10]{DBLP:series/txtcs/FominK10}),
and improves upon the naive $2^{\calO(|U|^2)} \cdot (|U|
+|\calF|)^{\calO(1)}$ brute-force algorithm.

Our next result is a linear vertex kernel when feedback edge set number (${\fes}$) is considered as the parameter.
It solves an open problem raised in~\cite{DBLP:journals/corr/abs-2011-14849,CGS21}.

\begin{restatable}{theorem}{locdomfesalgo}
\label{thm:LD-struct-fes}
\LD admits a kernel with $\calO(\fes)$ vertices and edges,
where \fes\ is the feedback edge set number of the input graph.
\end{restatable}

Our final result is about the incompressibility of both problems. 

\begin{restatable}{theorem}{incompressibility}
\label{thm:loc-dom-set-incompressibility}
Neither \LD\ nor \TCPB\ admits a polynomial compression of size
$\calO(n^{2 - \epsilon})$ for any $\epsilon > 0$, unless $\NP
\subseteq \coNP/poly$, where $n$ denotes the number of vertices and the
number of items and tests of the input, respectively.
\end{restatable} 

The reduction used to prove 
Theorem~\ref{thm:loc-dom-set-incompressibility} also provides an alternative proof that
\LD\ cannot be solved in time $2^{o(n)}$ (see~\cite{BIT20} for an earlier proof), and thus, in time
$2^{o(\vc)}$ (under the \ETH). Hence, the bound
of Theorem~\ref{thm:LD-struct-vc} is optimal up to a logarithmic factor.
The reduction used to prove 
Theorem~\ref{thm:loc-dom-set-incompressibility} also yields the following results.
First, it proves that \LD, parameterized by the vertex cover number of the 
input graph and the solution size, does not admit a polynomial kernel,
unless $\NP \subseteq \coNP/poly$.
Our reduction is arguably a simpler argument than the one
from~\cite{DBLP:journals/corr/abs-2011-14849,CGS21} to obtain the latter
result.
Second, it implies that \TCPB, parameterized by the number of items $|U|$
and the solution size $k$, does not admit a polynomial kernel,
unless $\NP \subseteq \coNP/poly$.
Additionally, it also implies that \TCPB, parameterized by the number $|\calF|$ of tests, does not admit a polynomial kernel,
unless $\NP \subseteq \coNP/poly$.
Since we can assume $k\leq |\calF|$ in any non-trivial instance, 
this improves upon the result in~\cite{GMY13},
that states that there is no polynomial kernel for the problem
when parameterized by $k$ alone.

\begin{figure}[!ht]%
\footnotesize{
  \centering
  \input{overview.tikz}%
}
\caption{Hasse diagram of graph parameters and associated results for {\sc Locating-Dominating Set}.
An edge from a lower parameter to a higher parameter indicates that the lower one is upper bounded by a function of the higher one. If the line is dashed, then the bound is exponential; otherwise, it is polynomial.
Colors correspond to the known FPT complexity status with respect to the highlighted  parameter: the upper half of the box represents the upper bound, and the lower half of the box represents the lower bound. By ``single-exponential'', ``slightly super-exponential'' and ``double-exponential'', we mean functions of the form $2^{\calO(p)}$, $2^{\calO(p\log p)}$ or $2^{\calO(p^2)}$, and $2^{2^{\calO(p)}}$, respectively. 
The parameters for which the running time is not known to be tight are thus striped.
The red circle in the upper-right corner means that \LD\ does not admit a polynomial kernel when parameterized by the marked parameter unless $\NP\subseteq \coNP/poly$; the yellow one means that a (tight) quadratic kernel exists, and the green one, that a linear kernel exists.
The bold borders highlight parameters that are covered in this paper. 
}
\label{fig:result-overview}
\vspace{-5mm}
\end{figure}
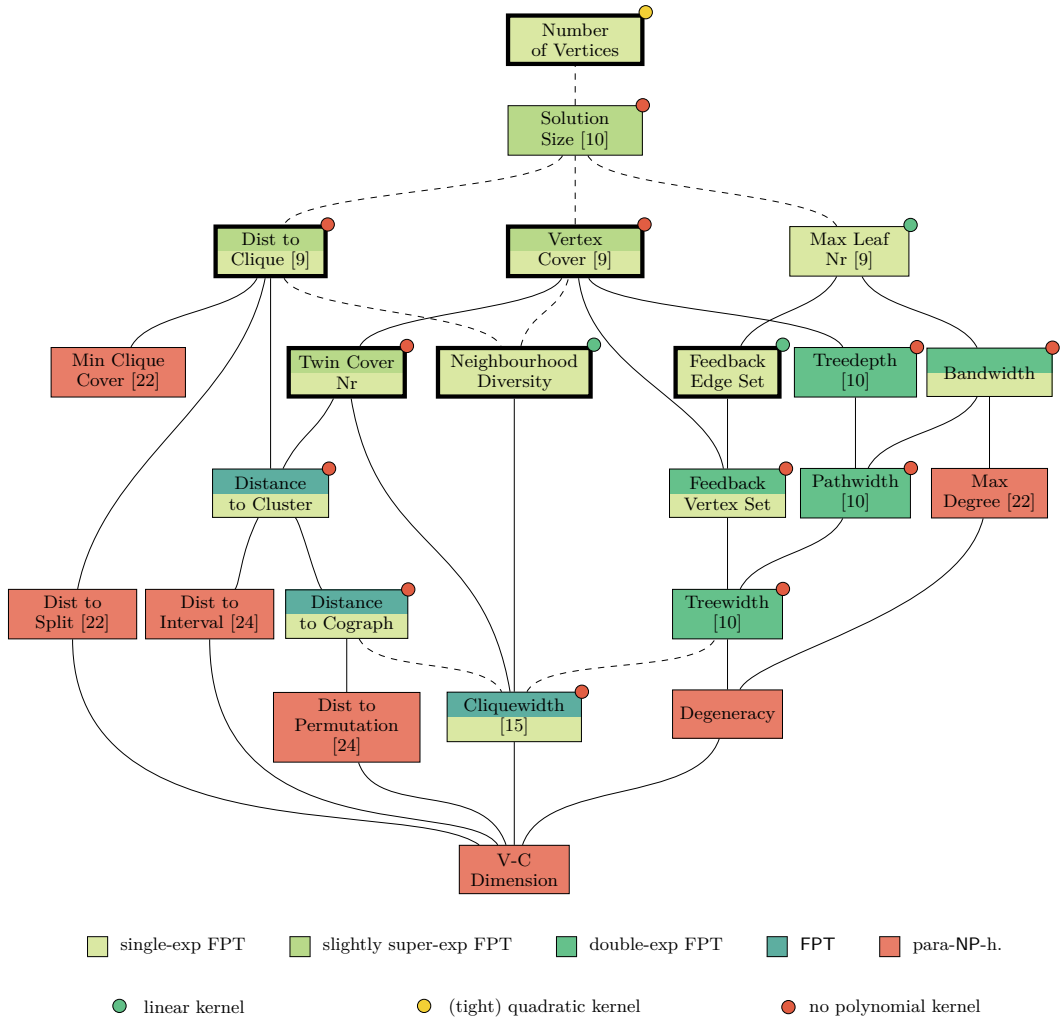


\subparagraph{Organization.}
We use the \LD\ problem to demonstrate key technical concepts, and explain how to adopt the arguments for \TCPB. 
We use the standard notations that are mentioned in 
Section~\ref{sec:prelims}.
Theorem~\ref{thm:LD-struct-vc} is proved in Section~\ref{sec:vc-fpt}. 
In the same section, we
show how to modify the techniques to obtain similar results
for other parameters like twin-cover number and 
distance to clique, and we observe that \LD 
admits a linear kernel when parameterized by neighbourhood diversity.
The proof of Theorem~\ref{thm:LD-struct-fes}
is given in Section~\ref{sec:fes-kernel} whereas
Theorem~\ref{thm:loc-dom-set-incompressibility}
is proved in Section~\ref{sec:incompressibility}.
We conclude the paper with some open problems in 
Section~\ref{sec:conclusion}.


%% file: overview.tikz
\definecolor{green}{RGB}{97, 191, 135}
\definecolor{red}{RGB}{226, 93, 66}
\definecolor{yellow}{RGB}{243, 208, 53}
\newcommand{\circgreen}{\raisebox{0.5pt}{\tikz{\node[draw,scale=0.4,circle,fill=yellow](){};}}}
\newcommand{\circred}{\raisebox{0.5pt}{\tikz{\node[draw,scale=0.4,circle,fill=yellow](){};}}}
\newcommand{\circyellow}{\raisebox{0.5pt}{\tikz{\node[draw,scale=0.4,circle,fill=yellow](){};}}}

\tikzstyle{mybox}=[align=center, rectangle, minimum height=.8cm,text width=2cm,draw]

\tikzset{
  circ/.style = {circle,draw,fill,inner sep=2.2pt}
}

\tikzstyle{paralb}=[mybox,preaction={path picture={\fill[red!80] (path picture bounding box.south west) rectangle (path picture bounding box.east);}}]
\tikzstyle{paraub}=[mybox,path picture={\fill[red!80] (path picture bounding box.north west) rectangle (path picture bounding box.east);}]
\tikzstyle{para}=[mybox,preaction={fill=red!80}]
\tikzstyle{fptlb}=[mybox,preaction={path picture={\fill[PineGreen!60] (path picture bounding box.south west) rectangle (path picture bounding box.east);}}]
\tikzstyle{fptub}=[mybox,path picture={\fill[PineGreen!60] (path picture bounding box.north west) rectangle (path picture bounding box.east);}]
\tikzstyle{fpt}=[mybox,preaction={fill=PineGreen!60}]
\tikzstyle{dexplb}=[mybox,preaction={path picture={\fill[Green!60] (path picture bounding box.south west) rectangle (path picture bounding box.east);}}]
\tikzstyle{dexpub}=[mybox,path picture={\fill[Green!60] (path picture bounding box.north west) rectangle (path picture bounding box.east);}]
\tikzstyle{dexp}=[mybox,preaction={fill=Green!60}]
\tikzstyle{sexplb}=[mybox,preaction={path picture={\fill[SpringGreen!60] (path picture bounding box.south west) rectangle (path picture bounding box.east);}}]
\tikzstyle{sexpub}=[mybox,path picture={\fill[SpringGreen!60] (path picture bounding box.north west) rectangle (path picture bounding box.east);}]
\tikzstyle{sexp}=[mybox,preaction={fill=SpringGreen!60}]
\tikzstyle{nsexplb}=[mybox,preaction={path picture={\fill[LimeGreen!60] (path picture bounding box.south west) rectangle (path picture bounding box.east);}}]
\tikzstyle{nsexpub}=[mybox,path picture={\fill[LimeGreen!60] (path picture bounding box.north west) rectangle (path picture bounding box.east);}]
\tikzstyle{nsexp}=[mybox,preaction={fill=LimeGreen!60}]

\tikzstyle{nottight}=[]
\tikzstyle{our}=[line width = 2pt]

\resizebox{\textwidth}{!}{
\begin{tikzpicture}[node distance=7mm]

	\node[sexplb,sexpub,our] (n) at (7.5, 9.5) {Number of Vertices};
	\node[nsexplb,nsexpub] (k) at (7.5, 8) {Solution Size \cite{DBLP:journals/corr/abs-2402-08346}};

	\node[sexplb,nsexpub,our] (vc) at (7.5, 6) {Vertex Cover \cite{CGS21}};
	\node[sexp, text width=1.75cm] (ml) at (12, 6)  {Max Leaf Nr \cite{CGS21}};
	\node[sexplb,nsexpub,our, text width=1.6cm] (dc) at (2.5, 6)  {Dist to \\ Clique \cite{CGS21}};

	\node[para] (mcc) at (0, 4) {Min Clique \\ Cover \cite{F15}};
 	\node[sexplb,nsexpub,our, text width=1.7cm] (tc) at (3.75, 4) {Twin Cover\\ Nr};
 	\node[sexplb,fptub,nottight, text width=1.7cm] (dcl) at (2.5, 2) {Distance\\ to Cluster};
 	\node[sexp,our,text width=2.3cm] (nd) at (6.5, 4) {Neighbourhood\\ Diversity};
 	\node[sexp,our, text width=1.5cm] (fes) at (10, 4) {Feedback\\ Edge Set};
	 \node[dexp, text width=1.8cm] (td) at (12.1, 4) {Treedepth \cite{DBLP:journals/corr/abs-2402-08346}};
 	\node[sexplb,dexpub,nottight,text width=1.85cm] (bw) at (14.3, 4) {Bandwidth};

 	\node[sexplb,fptub,nottight,text width=1.8cm] (dcg) at (3.75, 0) {Distance to Cograph};
 	\node[para,text width=1.9cm] (dsplit) at (-0.75, 0) {Dist to \\ Split \cite{F15}};
 	\node[para,text width=1.9cm] (dig) at (1.5, 0) {Dist to \\ Interval \cite{FMNPV17intervals2}};
 	\node[para,text width=2.2cm, below= of dcg,yshift=-1.8mm] (dperm) {Dist to \\ Permutation \cite{FMNPV17intervals2}};
 	\node[sexplb,dexpub,nottight, text width=1.7cm] (fvs) at (10, 2) {Feedback Vertex Set};
 	\node[dexp, text width=1.6cm] (pw) at (12.1, 2) {Pathwidth \cite{DBLP:journals/corr/abs-2402-08346}};
 	\node[para,text width=1.7cm] (mxd) at (14.3, 2) {Max Degree \cite{F15}};
 		
 	\node[dexp, text width=1.6cm] (tw) at (10, 0) {Treewidth \cite{DBLP:journals/corr/abs-2402-08346}};
 	 \node[sexplb,fptub,nottight,text width=2cm] (cw) at (6.5,-1.7) {Cliquewidth \\ \cite{C90}};
 	 \node[para,text width=1.6cm,below= of tw,yshift=-1.3mm] (deg) {Degeneracy};
 	 \node[para,text width=1.6cm,below= of cw,yshift=-1cm] (vcdim) {V-C Dimension};
 		
\newcommand{\nopolykernel}[1]{
 	\node[circ,fill=red, inner sep = 2.2pt] at ($#1+(0.0,0.0)$) {};
}
\newcommand{\polykernel}[1]{
 	\node[circ, fill=yellow, inner sep=2.2pt] at ($#1+(0.02,0.02)$) {};
}
\newcommand{\linearkernel}[1]{
 	\node[circ, fill=green, inner sep=2.2pt] at ($#1+(0.02,0.02)$) {};
}

\nopolykernel{(fvs.north east)}
\nopolykernel{(vc.north east)}
\nopolykernel{(tw.north east)}
\nopolykernel{(dc.north east)}
\nopolykernel{(td.north east)}
\nopolykernel{(pw.north east)}
\nopolykernel{(tc.north east)}
\nopolykernel{(dcl.north east)}
\nopolykernel{(dcg.north east)}
\nopolykernel{(cw.north east)}
\nopolykernel{(bw.north east)}
\nopolykernel{(k.north east)}
\linearkernel{(nd.north east)}
\linearkernel{(ml.north east)}
\linearkernel{(fes.north east)}
\polykernel{(n.north east)}
 		
 	\draw (dc) .. controls +(-0.5, -1) and +(0.5, 1) .. (mcc)
 	    (dc) -- (dcl)
 	    (vc) .. controls +(-0.5, -1) and +(0.5, 1) .. (tc)
 	    (tc) .. controls +(-0.5, -1) and +(0.5, 1) .. (dcl)
 	    (vc) .. controls +(0.5, -1) and +(-0.5, 1) .. (td)
 	    (vc) .. controls +(0.25, -2) and +(-0.25, 1.5) .. (fvs)
 	    (ml) .. controls +(-0.5, -1) and +(0.5, 1) .. (fes)
 	    (ml) .. controls +(0.5, -1) and +(-0.5, 1) .. (bw)
 	    (dcl) .. controls +(-0.5, -1) and +(0.5, 0.5) .. (dig)
 	    (dc) .. controls +(-0.5, -2.5) and +(0.5, 2.5) .. (dsplit)
 	    (td) -- (pw)
 	    (vcdim) -- (cw)
 	        (tc) .. controls +(0.5, -3) and +(-0.5, 3) .. (cw)
 	    (dperm) -- (dcg)
 	    (deg) -- (tw)
 	    (vcdim) .. controls +(0.5, 1.5) and +(-0.5, -1.5) .. (deg)
 	    (deg) .. controls +(0.5, 1) and +(-0.5, -2) .. (mxd)
 	    (pw) .. controls +(-0.5, -1) and +(0.5, 1) .. (tw)
 	    (cw) -- (nd)
 	    (bw) -- (mxd)
 	    (dsplit) .. controls +(0, -4) and +(-1.7, 1.2) .. (vcdim)
 	    (dig) .. controls +(0, -4) and +(-0.8, 1.2) .. (vcdim)
 	    (dperm) .. controls +(0.5, -1.5) and +(-0.5, 1.5) .. (vcdim)
 	    (fvs) -- (tw)
 	    (bw) .. controls +(-0.5, -1) and +(0.5, 1) .. (pw)
 	    (fes) -- (fvs)
 	    (dcl) .. controls +(0.5, -0.5) and +(-0.5, 0.5) .. (dcg)
		;

\draw[dashed]	(nd) .. controls +(0.25, 1) and +(-0.25, -1) .. (vc)
                (dc) .. controls +(0.5, -1) and +(-0.5, 1) .. (nd)
 		(dcg) .. controls +(0.5, -1) and +(-0.5, 1) .. (cw)
 	 	(cw) .. controls +(0.5, 1) and +(-0.5, -1) .. (tw)
                (k) .. controls +(-0.5, -1) and +(0.5, 1) .. (dc)
                (k) .. controls +(0.5, -1) and +(-0.5, 1) .. (ml)
		(vc) -- (k)
                (k) -- (n)
;
 	\node at (7, -5.5) (colors) {
	
	\fcolorbox{black}{SpringGreen!60}{\rule{0pt}{3pt}\rule{3pt}{0pt}} \ single-exp FPT~~~~~~
	\fcolorbox{black}{LimeGreen!60}{\rule{0pt}{3pt}\rule{3pt}{0pt}} \ slightly super-exp FPT~~~~~~
 	\fcolorbox{black}{Green!60}{\rule{0pt}{3pt}\rule{3pt}{0pt}} \ double-exp FPT~~~~~~
 	\fcolorbox{black}{PineGreen!60}{\rule{0pt}{3pt}\rule{3pt}{0pt}} \ \FPT~~~~~~
 	\fcolorbox{black}{red!80}{\rule{0pt}{3pt}\rule{3pt}{0pt}} \ para-\NP-h.};

\nopolykernel{(11,-6.5)}
\node at (12.75,-6.5) {no polynomial kernel};
\polykernel{(5,-6.5)}
\node at (7,-6.5) {(tight) quadratic kernel};
\linearkernel{(0,-6.5)}
\node at (1.25,-6.5) {linear kernel};

	\end{tikzpicture}
}

%% file: prelims.tex
\section{Preliminaries}
\label{sec:prelims}

For a positive integer $q$, we denote {the} set $\{1, 2, \dots, q\}$ by $[q]$.
We use $\mathbb{N}$ to denote the collection of all non-negative integers.

\subparagraph*{Graph theory}
We use standard graph-theoretic notation, and we refer the reader 
to~\cite{Diestel12} for any undefined notation. For an undirected graph $G$, 
sets $V(G)$ and $E(G)$ denote its set of vertices and edges, respectively.
We denote an edge with two endpoints $u, v$ as $uv$.
Unless otherwise specified, we use $n$ to denote the number of vertices in 
the input graph $G$ of the problem under consideration.
Two vertices $u, v$ in $V(G)$ are \emph{adjacent} if there is an edge $uv$ {in 
$G$}.
The \emph{open neighborhood} of a vertex $v$, denoted by $N_G(v)$, is the 
set of vertices adjacent to $v$.
The \emph{closed neighborhood} of a vertex $v$, denoted by $N_G[v]$, is 
the set $N_G(v) \cup \{v\}$.
We say that a vertex $u$ is a \emph{pendant vertex} if $|N_G(v)| = 1$.
We omit the subscript in the notation for neighborhood if the graph under 
consideration is clear.
For a subset $S$ of $V(G)$, we define $N[S] = \bigcup_{v \in S} N[v]$ and 
$N(S) = N[S] \setminus S$.
For a subset $S$ of $V(G)$, we denote the graph obtained by deleting $S$ 
from $G$ by $G - S$.
We denote the subgraph of $G$ induced on the set $S$ by $G[S]$.
Two vertices $u,v$ of a graph are called \emph{false twins} (respectively, \emph{true twins}) if $N(u) = N(v)$ (respectively, $N[u] = N[v]$). 
A pair of either false twins or true twins are also typically referred to as \emph{twins}.


\subparagraph*{Locating-Dominating Sets}
A subset of vertices $S$ in graph $G$ is called its \emph{dominating set}
if $N[S] = V(G)$.
A dominating set $S$ is said to be a \emph{locating-dominating set}
if for any two different vertices $u, v \in V(G) \setminus S$, we have $N(u) 
\cap S \neq N(v) \cap S $.
In this case, we say vertices $u$ and $v$ are
\emph{distinguished} by the set $S$.
We say a vertex $u$ is \emph{located} by set $S$
if for any vertex $v \in V(G) \setminus \{u\}$, $N(u) \cap S \neq N(v) \cap S$.
By extension,
a set $X$ is \emph{located} by $S$ if all vertices
in $X$ are located by $S$.

Next, we prove that it is safe to consider
the locating-dominating set that contains
all the vertices adjacent with pendant vertices (see also \cite[Lemma 5]{chakraborty2024n2boundlocatingdominatingsetssubcubic}).

\begin{observation}
\label{obs:nbr-of-pendant-vertex-in-sol}
If $S$ is a locating-dominating set of a graph $G$,
then there exists a locating-dominating set $S'$ of $G$
such that $|S'| \le |S|$ and $S'$ contains all vertices that are adjacent to the pendant vertices (i.e. vertices of degree~1) in $G$.
\end{observation}
\begin{claimproof}
Let $u$ be a pendant vertex which is adjacent to a vertex $v$ of $G$. We now look for a locating dominating set $S'$ of $G$ such that $|S'| \le |S|$ and contains
the vertex $v$.
As $S$ is a (locating) dominating set,  we have $\{u, v\} \cap S \neq 
\emptyset$. If $v \in S$, then take $S' = S$.
Therefore, let us assume that $u \in S$ and $v \not\in S$.
Define $S' = (S \cup \{v\}) \setminus \{u\}$.
It is easy to see that $S'$ is a dominating set.
If $S'$ is not a locating-dominating set, then there exists $w$, apart from 
$u$,
in the neighbourhood of $v$ such that both $u$ and $w$ are adjacent
with \emph{only} $v$ in $S'$.
As $u$ is a pendant vertex and $v$ its unique neighbour, $w$ is not adjacent 
with $u$.
Hence, $w$ was not adjacent with any vertex in $S' \setminus \{v\} = S 
\setminus \{u\}$.
This, however, contradicts the fact that $S$ is a (locating) dominating set.
Hence, $S'$ is a locating-dominating set and $|S'| = |S|$. Thus, the result follows from repeating this argument for every vertex of $G$ that is adjacent to a pendant vertex.
\end{claimproof}


\subparagraph*{Parameterized complexity}
\label{prelim:pc}
An instance of a parameterized problem $\Pi$ {consists} of an input $I$, 
which is an input of the non-parameterized version of the problem, and an 
integer $k$, which is called the \emph{parameter}.
Formally, $\Pi \subseteq \Sigma^* \times \mathbb{N}$.
A problem $\Pi$ is said to be \emph{fixed-parameter tractable}, or \FPT, if 
given an instance $(I,k)$ of $\Pi$, we can decide whether  $(I,k)$ is a 
\yes-instance 
of $\Pi$ in  time $f(k)\cdot |I|^{\calO(1)}$.
Here, {$f: \mathbb{N} \mapsto \mathbb{N}$} is some computable function 
{depending} only on $k$.
%

A parameterized problem $\Pi$ is said to admit a {\em kernelization} if given 
an instance $(I, k)$ of $\Pi$, there is an algorithm that runs in time 
polynomial in $|I| + k$ and constructs an instance $(I', k')$ of $\Pi$ such that 
{(i)} $(I, k) \in \Pi$ if and only if $(I', k')\in \Pi$, and {(ii)} $|I'| + k' \leq 
g(k)$ for some computable function $g: \mathbb{N} \mapsto \mathbb{N}$ 
depending only on $k$.
If $g(\cdot)$ is a polynomial function, then $\Pi$ is said to admit a 
{\em polynomial} kernelization.
In some cases, it is possible that a parameterized problem $\Pi$ is 
preprocessed into an equivalent instance of a different problem satisfying 
some special properties.
Such preprocessing algorithms are also well-known in the literature.
A {\em polynomial compression} of a parameterized problem $\Pi$ into a 
problem $\Psi \subseteq \Sigma^*$ is an algorithm that given an instance 
$(I, k)$ of $\Pi$, runs in polynomial in $|I| + k$ time, and outputs an instance 
$I'$ of $\Psi$ such that {(i)} $(I, k) \in \Pi$ if and only if $I' \in \Psi$, and 
{(ii)} $|I'| \leq f(k)$ for a polynomial function $f: \mathbb{N} \mapsto 
\mathbb{N}$.
If $|\Sigma| = 2$, then the function $f(\cdot)$ is called the {\em bitsize} of 
the compression.

A \emph{reduction rule} is a polynomial time  algorithm that takes as input 
an instance of a problem and outputs another, usually reduced, instance.
A reduction rule said to be \emph{applicable} on an instance if the output 
{instance} and input instance are different.
A reduction rule is \emph{safe} if the input instance is a \yes-instance if and 
only if the output instance is a \yes-instance.
For {a} detailed introduction to parameterized complexity and related 
terminologies, we refer the reader to the recent books by Cygan et 
al.~\cite{cygan2015parameterized} and Fomin et 
al.~\cite{fomin2019kernelization}.

%% file: vc-fpt.tex
\section{Parameterization by Vertex Cover Number}
\label{sec:vc-fpt}

We describe an algorithm to prove Theorem~\ref{thm:LD-struct-vc},
for \LD, i.e., to prove that it admits an algorithm running in time 
$2^{\calO(\vc\log\vc)} \cdot n^{\calO(1)}$, where 
\vc\ is the vertex cover number of the input graph.
Our algorithm is based on a reduction to a dynamic programming
scheme for a generalized partition refinement problem, 
that can also be used to \TCPB.
We start with the following reduction rule 
which is applicable in polynomial time.

\begin{reduction rule}\label{reduc:twins}
Let $(G,k)$ be an instance of \LD. 
If there exist three vertices
$u,v,x$ of $G$ such that any two of $u,v,x$ are twins, then delete $x$
from $G$ and decrease $k$ by one.
\end{reduction rule}

\begin{lemma}
\label{lem:reduc-twins}
Reduction Rule~\ref{reduc:twins} is correct and can be applied in time 
$\calO(n+m)$.
\end{lemma}
\begin{proof}
We show that $(G,k)$ is a \yes-instance of \LD if 
and only if $(G - \{x\},k-1)$ is a \yes-instance of \LD.
Slater proved that for any set $S$ of vertices of a graph $G$ such that any two vertices in $S$ are twins, any locating-dominating set contains at least $|S|-1$ vertices of  $S$~\cite{slater1988dominating}.
Hence, any locating-dominating set must contain at least 
two vertices in $\{u, v, x\}$. We can assume, without loss of generality, that 
any resolving set contains both $u$ and $x$.
Hence, any pair of vertices in $V(G) \setminus \{u, x\}$ that 
is located by $x$ is also located by $u$, moreover, any vertex in $V(G) 
\setminus \{u, x\}$ that is dominated by $x$ is also dominated by $u$.
Thus, if $S$ is a locating-dominating set of $G$, then
$S \setminus \{x\}$ is a locating-dominating set of $G - \{x\}$. This implies 
that if $(G,k)$ is a \yes-instance, then $(G - \{x\},k-1)$ is a \yes-instance.
The correctness of the reverse direction follows from the fact
that we can add $x$ into a locating-dominating set of $G - \{x\}$
to obtain one of $G$.
Since detecting twins in a graph can be done in time $\calO(n+m)$~\cite{HPV98}, 
the running time follows.
\end{proof}

Our algorithm starts by finding a minimum vertex cover, say $U$, of $G$ by an algorithm in time $1.2528^{\vc} \cdot n^{\calO(1)}$ \cite{DBLP:conf/stacs/0001N24}.
Then, $\vc = |U|$. Moreover, let 
$R = V \setminus U$ be the corresponding independent set in $G$. 
The algorithm first applies 
Reduction Rule~\ref{reduc:twins} exhaustively to reduce every 
twin-class to size at most~$2$. 
For the simplicity of notations, we continue to call the 
reduced instance of \LD as $(G,k)$.

Consider an optimal but hypothetical locating-dominating set $L$ of $G$.
The algorithm constructs a partial solution $Y_L$ and, in subsequent
steps, expands it to obtain $L$.
The algorithm initializes $Y_L$ as follows: for all pairs $u,v$ of 
twins in $G$, it adds one of them in $Y_L$. 
Slater proved that for any set $S$ of vertices of a graph $G$ such that any two vertices in $S$ are twins, any locating-dominating set contains at least $|S|-1$ vertices of  $S$~\cite{slater1988dominating}.
Hence, it is safe to assume that 
all the vertices in $Y_L$ are present in any locating-dominating set.
Next, the algorithm guesses the intersection of $L$ with $U$.
Formally, it iterates over all the subsets of $U$, and for each such set,
say $X_L$, it computes a locating-dominating set of appropriate
size that contains all the vertices in $X_L$ and no vertex in 
$U \setminus X_L$.
Consider a subset $X_L$ of $U$ and define $B = U \setminus X_L$
and $R' = R \setminus (N(X_L) \cup Y_L)$. 
As $L$ is also a dominating set, it is safe to assume that $R' \subseteq L$.
The algorithm updates $Y_L$ to include $X_L \cup R'$.

\begin{figure}[t]
\begin{center}
\includegraphics[scale=0.5]{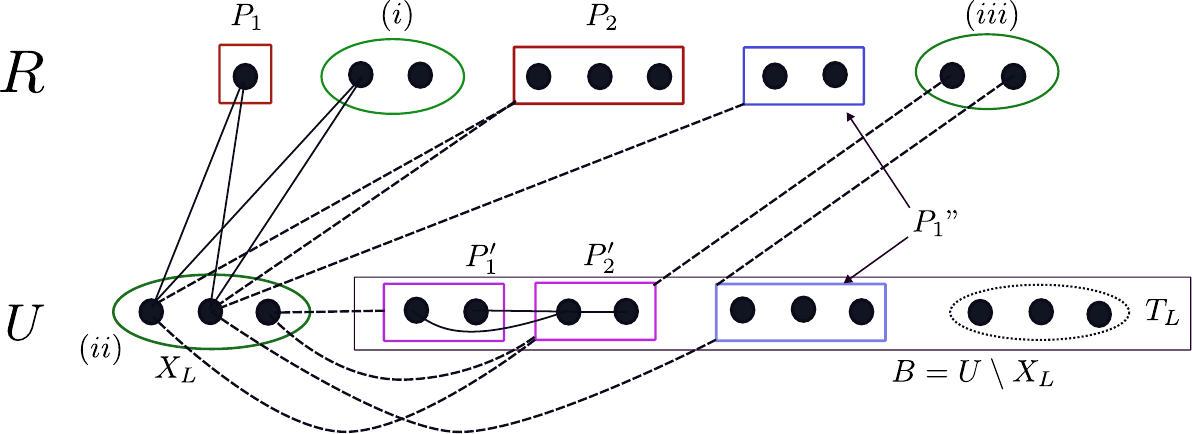}
\end{center}
\caption{An instance $(G,k)$ of \LD.
Set $U$ is a minimum-size vertex cover of $G$.
The dotted edge denotes that a vertex is adjacent with all the 
vertices in the set.
For the sake of brevity, we do not show all the edges. 
The vertices in green ellipses are part of solution because
of $(i)$ being a part of false-twins, $(ii)$ guessed intersection
with $U$, and $(iii)$ the requirement of solution to be a dominating
set. 
Note that the vertices $T_L$ are not dominated by the partial solution $Y_L$.
Parts $P_1, P_2, P'_1, P'_2,$ and $P''_1$ are parts of partition of 
$R \cup B$ induced by $Y_L$.
 \label{fig:loc-dom-set-vc}}
 \vspace{-3mm}
\end{figure}

At this stage, $Y_L$ dominates all the vertices in $R$.
However, it may not dominate all the vertices in $B$.
The remaining (to be chosen) vertices in $L$ are part of $R$ and are
responsible for dominating the remaining vertices in $B$ and to locate
all the vertices in $R \cup B$. See Figure~\ref{fig:loc-dom-set-vc}
for an illustration.  As the remaining solution, i.e., $L \setminus
Y_L$, does not intersect $B$, it is safe to ignore the edges both of
whose endpoints are in $B$. The vertices in $Y_L$ induce a partition
of the vertices of the remaining graph, according to their
neighborhood in $Y_L$. We can redefine the objective of selecting the remaining
vertices in a locating-dominating set as to \emph{refine} this
partition such that each part contains exactly one vertex. Partition
refinement is a classic concept in algorithms,
see~\cite{HPV98}. However in our case, the partition is not
standalone, as it is induced by a solution set.
To formalize this intuition, we introduce the following notation.

\begin{definition}[Partition Induced by $C$ and Refinement]
For a subset $C \subseteq V(G)$, define an equivalence
relation $\sim_C$ on 
$V' \subseteq V(G) \setminus C$ as follows:
for any pair $u, v \in V'$, $u \sim_C v$ if and only if 
$N(u) \cap C = N(v) \cap C$.
Then, for any set $V \subseteq V(G)$, the partition of $V$ induced by $C$, denoted by $\calP(C)$, is the defined as follows:
$$\calP(C) = \{\{c\} \mid c \in V \cap C\} \cup \{S \mid S \text{ is 
an equivalence class of $\sim_C$ defined on } V \setminus C \}.$$
Moreover, for two partitions $\calP$ and $\mathcal{Q}$ of $V$, 
a \emph{refinement} of $\calP$ by $\q$, denoted by $\calP \Cap \q$,
is the partition defined as $\{P \cap Q : P \in \calP, Q \in \q\}$.
\end{definition} 

Suppose $V = V(G)$ and $\calP$ and $\q$ are two partitions of $V(G)$,
then $\calP \Cap \q$ is also a partition of $V(G)$.
In addition, it can be checked that, if $C_1, C_2 \subseteq V(G)$, then $\calP(C_1 \cup C_2) = \calP(C_1) \Cap \calP(C_2)$ on any subset $V \subseteq V(G)$.
We say a partition $\calP$ is the \emph{identity partition} 
if every part of $\calP$ is a singleton set.
This implies that a set $C \subseteq V(G)$ for any graph $G$ has the property $N(u) \cap C \ne N(v) \cap C$ for all distinct $u,v \in V(G) \setminus C$ if and only if the partition $\calP(C)$ of $V(G)$ is identity.
With these definitions, we define the following auxiliary problem.

\defproblem{\textsc{Annotated Red-Blue Partition Refinement}}{A bipartite
graph $G$ with bipartition $\langle R, B\rangle$ of $V(G)$;
a partition $\q$ of $R \cup B$;
a collection of forced solution vertices $C_0 \subseteq R$;
a collection of vertices $T_L \subseteq B$ that needs to be dominated;
and an integer $\lambda$.}{Does there exist a set $C$ of size at most $\lambda$ 
such that $C_0 \subseteq C \subseteq R$, 
$C$ dominates $T_L$, and 
$\q \Cap \calP(C)$ is the identity partition of $R \cup B$?}

Suppose there is an algorithm $\calA$ that solves \textsc{Annotated Red-Blue Partition Refinement} in time $f(|B|) \cdot (|R| + |B|)^{\calO(1)}$.
Then, there is an algorithm that solves \LD
in time $2^{\calO(\vc)} \cdot f(\vc) \cdot n^{\calO(1)}$.
Consider the algorithm described so far in this subsection.
The algorithm then calls $\calA$ as a subroutine
with the bipartite graph $\langle R, B\rangle$ obtained from $G$, where $B = U \setminus X_L$ and $R = V \setminus U$ (notice that, since the hypothetical solution $L$ has no vertices from $B$, the edges of $G$ with both endpoints in $B$ are irrelevant to the locating property of $L$).
It sets $\q$ as the partition of $R \cup B$ induced by $Y_L$,
$C_0 = Y_L \cap R$, $T_L = B \setminus N[Y_L]$,
and $\lambda = k - |X_L|$.
Note that $T_L$ is the collection of vertices that
are \emph{not} dominated by the partial solution $Y_L$.
The correctness of the algorithm follows from the 
correctness of $\calA$ and the fact that 
for two sets $C_1, C_2 \subseteq V(G)$,
$C_1 \cup C_2$ is a locating-dominating set of $G$
if and only if $C_1 \cup C_2$ is a dominating set and $\calP(C_1 \cup C_2) = \calP(C_1) \Cap \calP(C_2)$ is the identity 
partition of $V(G)$. 
Hence, it suffices to prove the following lemma. 

\begin{lemma}
\label{lemma:partition-refinement-fpt}
There is an algorithm that solves \textsc{Annotated Red-Blue Partition Refinement} in time $2^{\calO(|B|\log|B|)} \cdot (|R| + |B|)^{\calO(1)}$.
\end{lemma}

The remainder of the section is devoted to prove Lemma~\ref{lemma:partition-refinement-fpt}. 
We present the algorithm that can roughly be divided into three parts.
In the first part, the algorithm processes the partition 
$\q$ of $R \cup B$ with the goal of reaching to
a refinement of partition $\q$ such that every part is completely
contained either in $R$ or in $B$.
Then, we introduce some terms to obtain a `set-cover' type 
dynamic programming.
We also introduce three conditions that restrict the number 
of dynamic programming states we need to consider to 
$2^{\calO(|B|\log(|B|))} \cdot |R|$.
Finally, we state the dynamic programming procedure,
prove its correctness and argue about its running time. 

\begin{figure}[t]
\centering
\includegraphics[scale=0.5]{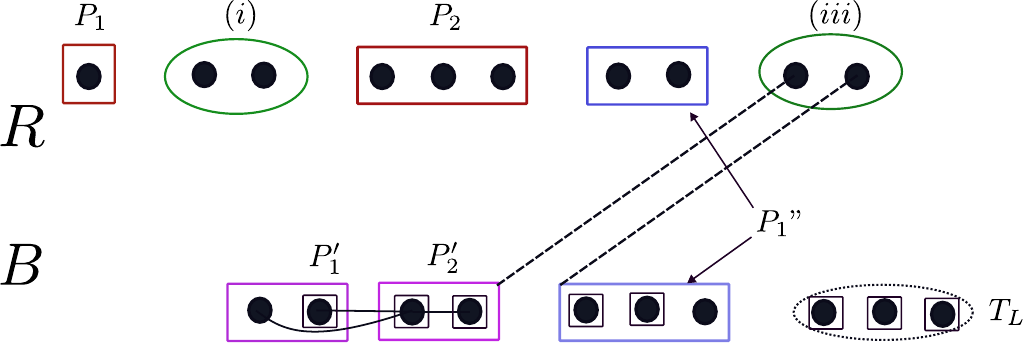}
\caption{Instance of the \textsc{Annotated Red-Blue Partition Refinement} problem.
Vertices in green ellipse denotes set $C_0$ whereas dotted ellipse
denote vertices in $T_L$, and
the partition $\q = \{P_1, P_2, P'_1, P'_2, P''_1\}$.
The vertices in the rectangles denote vertices in $T^{\circ}_L$.
\label{fig:instance-of-red-blue-partition-refinement}}
\vspace{-3mm}
\end{figure}

\subparagraph*{Pre-processing the Partition.}
Consider the partition $\calP(C_0)$ of $R \cup B$ induced by $C_0$,
and define $\calP_0 = \q \Cap \calP(C_0)$.
We classify the parts in $\calP_0$ into three classes 
depending on whether they intersect with $R$, $B$, or both.
Let $P_1, P_2, \ldots , P_t$ be an arbitrary but fixed order on 
the parts of $\calP_0$ that are completely in $R$.
Similarly, let $P'_1, P'_2, \ldots , P'_{t'}$ be an arbitrary but fixed order
on the parts that are completely in $B$. 
Also, let $P''_1, P''_2, \ldots , P''_{t''}$ be the collection 
of parts that intersect $R$ as well as $B$.
Formally, we have $P_j \subseteq R$ for any $j \in [t]$, 
$P'_{j'} \subseteq B$ for any $j' \in [t']$, and
$P''_{j''} \cap R \neq \emptyset \neq P''_{j''} \cap B$ for any $j'' \in [t'']$.
See Figure~\ref{fig:instance-of-red-blue-partition-refinement} for an illustration.

Recall that $T_L$ is the collection of vertices in $B$ that are required
to be dominated.
The algorithm first expands this set so that it is precisely
the collection of vertices that need to be dominated by the solution $C$.
In other words, at present, the required condition is $T_L \subseteq N(C)$,
whereas, after the expansion, the condition is $T_L = N(C)$.
Towards this, it first expands $T_L$ to include $N(C_0)$.
It then uses the property that the final partition needs to be the 
identity partition to add some more vertices in $T_L$.

Consider parts $P'_{j'}$ or $P''_{j''}$ of $\calP_0$.
Suppose that $P'_{j'}$ (respectively, $P''_{j''} \cap B$) contains two 
vertices that are \emph{not} 
dominated by the final solution $C$, then these two vertices 
would not be separated by it and hence would remain in one part, a contradiction. 
Therefore, any feasible solution $C$ needs to dominate all
vertices but one in $P'_{j'}$ (respectively in $P''_{j''} \cap B$).
\begin{itemize}[nolistsep]
\item 
The algorithm modifies $T_L$ to include $N(C_0)$.
Then, it iterates over all the subsets $T^{\circ}_L$ of
$B \setminus T_{L}$ such that for any $j' \in [t']$ and $j'' \in [t'']$,
$|P'_{j'} \cap (T^{\circ}_{L} \cup T_{L})| \le 1$ and $|(P''_{j''} \cap B) \cap (T^{\circ}_{L} \cup T_{L})| \le 1$.
\end{itemize}


Consider a part $P_j$ of $\calP_0$, which is completely contained in $R$.
As $R$ is an independent set and the final solution $C$ 
is not allowed to include any vertex in $B$, 
for any part $P_j$, there is at most one vertex outside $C$,
i.e., $|P_j \setminus C| \le 1$.
However, unlike the previous step, the algorithm 
cannot enumerate all the required subsets of $R$ in the desired time.
Nonetheless, it uses this property to safely perform
the following sanity checks and modifications.
\begin{itemize}[nolistsep]
\item
Consider parts $P_{j}$ or $P''_{j''}$ of $\calP_0$. 
Suppose there are two vertices in $P_{j}$ (respectively $P''_{j''} \cap R$) 
that are adjacent with some vertices in $B \setminus T_L$.
Then, discard this guess of $T^{\circ}_L$ and move to the next subset. 
\item Consider parts $P_{j}$ or $P''_{j''}$ of $\calP_0$. 
Suppose there is a unique vertex, say $r$, in $P_{j}$ 
(respectively, in $P''_{j''}\cap R$) that is 
adjacent with some vertex in $B \setminus T_L$.
Then, modify $C_0$ to include all the vertices in $P_{j}\setminus \{r\}$ 
or  in $(P''_{j''} \cup R)\setminus \{r\}$.
\end{itemize}

Consider a part $P''_{j''}$ of $\calP_0$ and 
suppose $|(P''_{j''} \cap B) \setminus T_L| = 0$.
Alternately, every vertex in $P''_{j''} \cap B$ is adjacent to 
some vertex in the final solution $C$.
As $R$ is an independent set and $C \subseteq R$,
this domination condition already refines the partition $P''_{j''}$,
i.e. every vertex in $(P''_{j''} \cap B)$ needs to be adjacent 
with some vertex in $C$ whereas no vertex in 
$(P''_{j''} \cap R)$ can be adjacent with $C$.
This justifies the following modification.
\begin{itemize}[nolistsep]
\item For a part $P''_{j''}$ of $\calP_0$ such that 
$|(P''_{j''} \cap B) \setminus T_L| = 0$, the algorithm modifies 
the input partition $\calP_0$ 
by removing $P''_{j''}$ and adding two new parts 
$P_{j} := P''_{j''} \cap R$ and $P'_{j'} := P''_{j''} \cap B$.
\end{itemize}
Consider a part $P''_{j''}$ of $\calP_0$ and 
suppose $|(P''_{j''} \cap B) \setminus T_L| = 1$.
Let $b$ be the unique vertex in $(P''_{j''} \cap B) \setminus T_L$.
Suppose there is a (unique) vertex, say $r$, in $(P''_{j''} \cap R)$ that 
is not part of the final solution $C$.
However, this implies that both $b$ and $r$ are in the same part,
a contradiction.
\begin{itemize}[nolistsep]
\item For part $P''_{j''}$ of $\calP_0$ such that 
$|(P''_{j''} \cap B) \setminus T_L| = 1$, 
the algorithm includes all the vertices $(P''_{j''} \cap R)$ into 
$C_0$ (which, by definition, modifies $\calP_0)$.
Also, it removes $P''_{j''}$ from $\calP_0$ and adds $P'_{j'} := P''_{j''} \cap B$.  
\end{itemize}
These modifications ensure that any part of $\calP_0$ is
either completely contained in $R$ or completely contained in $B$.

\subparagraph*{Setting up the Dynamic Programming.}
Define $\ell = |R \setminus C_0|$ and 
suppose $\{r_1, r_2, \dots, r_{\ell}\}$ be the 
reordering of vertices in $R \setminus C_0$ such that
any part in $P_j$ contains the consecutive elements in the order.
Formally, for any $i_1 < i_2 < i_3 \in \{1, 2, \dots, |R \setminus C_0|\}$
and $t^{\circ} \in \{1, 2, \dots, t\}$,
if $r_{i_1} \in P_{t^{\circ}}$ and $r_{i_3} \in P_{t^{\circ}}$,
then $r_{i_2} \in P_{t^{\circ}}$.
Define a function $\pi: \{1, 2, \dots, |R \setminus C_0|\} 
\mapsto \{1, 2, \dots, t\}$
such that $\pi(r_i) = t^{\circ}$ if and only if 
vertex $r_i$ is in part $P_{t^{\circ}}$.

For the remaining part, the algorithm relies on a `set cover' type
dynamic programming.\footnote{See \cite[Theorem
  3.10]{DBLP:series/txtcs/FominK10} for a simple $\calO(mn2^n)$
dynamic programming scheme for \textsc{Set Cover} on a universe of
size $n$ and $m$ sets.}  For every integer $i \in [0, \ell]$, define
$R_i = \{r_1, r_2, \dots, r_i\}$.  Then, in a `set cover'-type dynamic
programming, tries to find the optimum size of a subset of $R_i$ that
dominates all the vertices in $S$, which is a subset of $T$. However,
two different optimum solutions in $R_i$ will induce different
partitions of $R \cup B$.  To accommodate this, we define the notion
of \emph{valid tuples}. For an integer $i \in [0,\ell]$ and a subset
$S \subseteq T$, we say a tuple $(i,\calP,S)$ is a \emph{valid tuple} if
$\calP$ is a refinement of $\calP_0$ and it satisfies the three
properties mentioned below. Note that these properties are the
consequences of the fact that we require $\calP$ to be such that, for
any (partial) solution $C$ with vertices upto $r_i$ in the ordering of
$R \setminus C_0$, we can have $\calP(C) = \calP$ and $N(C) = S$.
\begin{enumerate}[nolistsep]
\item For any $i \in [0,\ell]$, the partition $\calP$ restricted 
to any $P_j$, where $j \in [\pi(i)-1]$, is the identity partition of $P_j$.
This is because, if on the contrary, $\calP$ restricted to $P_j$ for 
some $j \in [\pi(i)-1]$ is not the identity partition, it would imply 
that there  exists a part $P$, say, of $\calP$ with $|P| \geq 2$ 
which cannot be refined by picking vertices of $R$ 
in the final solution $C$ (upto $r_\ell$), as $R$ is an independent set.
\item There may be at most one vertex $r$, say, in $P_{\pi(i)} \cap R_i$ 
which is not yet picked in $C$. This is because if two vertices 
$r',r''$ of $P_{\pi(i)} \cap R_i$ are not picked in $C$, 
the final solution $C$ cannot induce the identity partition, 
since $\{r'\}, \{r''\} \notin \calP(C)$. 
Thus, since for any $r' \in P_{\pi(i)} \cap R_i \setminus \{r\}$, 
we have $\{r'\} \in \calP(C)$, we therefore require the partition 
$P$ of $\calP$ to be a singleton set if $P \cap P_{\pi(i)} \cap R_i \neq \emptyset$ and $r \notin P$.
Moreover, since $(\{r\} \cup P_{\pi(i)} \setminus R_i) \in \calP(C)$,
we require that for $r \in P$, the part $P$ must be $(\{r\} \cup P_{\pi(i)} 
\setminus R_i) \in \calP(C)$.
\item Since any (partial) solution upto $r_i$ does not refine 
any of the parts $P_{\pi(i)+1}, \ldots , P_t$, for every part $P \in \calP$ 
that does not intersect $P_1 \cup P_2 \cup \ldots \cup P_{\pi(i)}$, $P$
must be of the form $P_x$, where $x \in [\pi(i)+1,t]$.
\end{enumerate}

From the above three properties of how the refinements 
$\calP$ of $\calP_0$ can be, we find that $\calP$ restricted 
to $B$ can be any partition on $B$ 
(which can be at most $\calO(2^{\calO(|B| \log |B|)})$ many) 
and the sets $P \in \calP$ containing the vertex $r \in P_{\pi(i)}$ 
can be as many as the number of vertices in $P_{\pi(i)} \cap R_i$, 
that is, at most $|R|$ many. 
Let $\calX$ be the collection of all the valid tuples.
Then, for any $i \in [\ell]$ and any $S \subseteq T$, the number of 
partitions $\calP$ for which $(i,\calP,S) \in \x$ is 
$\bigo(2^{\bigo(|B| \log |B|)} \cdot |R|)$.

\subparagraph*{Dynamic Programming.}
For every valid tuple $(i,\calP,S) \in \mathcal{X}$, we define
\begin{center}
\begin{tabular}{lcl}
$\lc[i,\calP,S]$ & = & the minimum cardinality of a set $C$ that is \emph{compatible} with the\\
& &  valid triple $(i,\calP,S)$, i.e., $(i)\ C \subseteq R_i\ (ii)\ \calP_0 \Cap \calP(C) = \calP\
\&\ (iii)\ N(C) = S$.
\end{tabular}
\end{center}

If no such $(i,\calP,S)$-compatible set exists, then we let the value of 
$\lc[i,\calP,S]$ to be $\infty$. 
Moreover, any $(i,\calP,S)$-compatible set $C$ of minimum cardinality is called a \emph{minimum $(i,\calP,S)$-compatible set}.
The quantities $\lc[i,\calP,S]$ for all $(i,\calP,S) \in \x$ are updated 
inductively. 
Then finally, the quantity $\lc[\ell, \mathcal{I}(R \cup B),T]$ 
gives us the required output of the problem. 
To start with, we define the quantity $\lc[0,\calP_0,N(C_0)] = |C_0|$ 
and set $\lc[i,\calP,S] = \infty$ for each triple $(i,\calP,S) \in \mathcal{X}$ 
such that $(i,\calP,S) \neq (0,\calP_0,N(C_0))$. 
Then, the value of each $\lc[i,\calP,S]$ updates by the following dynamic programming formula.
\begin{flalign}
\label{eq_dp}
\lc[i,\calP,S] = \min \begin{cases}
				  \lc[i-1,\calP,S],\\
				  1+ \min\limits_{\substack{\calP' \Cap \calP(r_i) = \calP,\\ S' \cup N(r_i) = S}}\lc[i-1,\calP',S'].
				  \end{cases}
\end{flalign}
\subparagraph*{Proof of Correctness.}
We prove by induction on $i$ that the above formula is correct. 
In other words, we show that, for any $i \in [\ell]$, 
if the values of $\lc[j,\calP',S']$ are correctly calculated for 
all triples $(j,\calP',S') \in \mathcal{X}$ with $j \in [0,i-1]$, 
then for any couple $(\calP,S)$, the above equality holds.
Before this, we prove some technical properties.

\begin{claim} \label{claim_gamma(iPS)_finite}
Let $(i,\calP,S) \in \x$. Then, the following hold.
\begin{itemize}[nolistsep]
\item The triple $(i-1,\calP,S) \in \x$ as well. Moreover, if $C$ is a $(i-1,\calP,S)$-compatible set, then $C$ is also a $(i,\calP,S)$-compatible set. In particular, $\lc[i,\calP,S] \leq \lc[i-1,\calP,S]$.
\item Let $(i-1,\calP',S') \in \x$ such that $\calP' \Cap \calP(r_i) = \calP$ and $S' \cup N(r_i) = S$. If $C'$ is a $(i-1,\calP',S')$-compatible set, then $C = C' \cup \{r_i\}$ is a $(i,\calP,S)$-compatible set. In particular, $\lc[i,\calP,S] \leq 1+\lc[i-1,\calP',S']$.
\end{itemize}
\end{claim}

\begin{proof}
We first show that $(i-1,\calP,S) \in \x$. 
So, let $P \in \calP$ be such that $P \cap R_{i-1} \neq \emptyset$ 
and $|P| \geq 2$. 
It implies that $P \cap R_i \neq \emptyset$ as well and 
hence, $P \cap P_j = \emptyset$ for all $j \in [0,\pi(i)-1]$ 
and $P \cap P_{\pi(i)}$ is singleton. 
If $\pi(i-1) < \pi(i)$, then $\pi(i-1) = \pi(i)-1$ and 
hence, $P \cap P_j = \emptyset$ for all $j \in [0,\pi(i-1)]$. 
In other words, $P \cap R_{i-1} = \emptyset$, 
a contradiction to our assumption. 
Therefore, we must have $\pi(i-1) = \pi(i)$ and thus, 
clearly, $(i-1,\calP,S) \in \x$. 
Now, let $C$ is a $(i-1,\calP,S)$-compatible set. 
Then clearly $C$ is also a $(i,\calP,S)$-compatible set. Then, the inequality follows immediately.

Now, we prove the second point. 
Let $C'$ be a $(i-1,\calP',S')$-compatible set. Then, we have the following.
\begin{itemize}[nolistsep]
\item $C = C' \cup \{r_i\} \subset R_i$;
\item $\calP_0 \Cap \calP(C) = \calP_0 \Cap \calP(C') \Cap \calP(r_i) = \calP' \cap \calP(r_i) = \calP$; and
\item $N(C) = N(C') \cup N(r_i) = S' \cup N(r_i) = S$.
\end{itemize}
This implies that the set $C$ is a $(i,\calP,S)$-compatible set. Now, if $\lc[i-1,\calP',S'] = \infty$, then, clearly, $\lc[i,\calP,S] \leq 1+\infty = 1+\lc[i-1,\calP,S]$. Lastly, if $\lc[i-1,\calP',S'] < \infty$ and $C'$ is a minimum $(i-1,\calP',S')$-compatible set, since $C$ is $(i,\calP,S)$-compatible, we have $\lc[i,\calP,S] \leq |C| = 1+|C'| = 1+\lc[i-1,\calP',S']$.
\end{proof}

We now prove by induction on $i$ that Equation~(\ref{eq_dp}) is correct. 
In other words, we show that, for any $i \in [\ell]$, if the values of 
$\lc[j,\calP',S']$ are correctly calculated for all triples 
$(j,\calP',S') \in \mathcal{X}$ with $j \in [0,i-1]$, 
then for any couple $(\calP,S)$, the above equality holds.
Let us therefore assume that, for all triples $(j,\calP',S')$ with 
$j \in [0,i-1]$, the values of $\lc[j,\calP',S']$ are correctly calculated.

To begin with, we show that if no $(i,\calP,S)$-compatible set exists, then Equation~(\ref{eq_dp}) must compute $\infty$ as the value of $\lc[i,\calP,S]$. In other words, we prove that the value of the RHS in Equation~(\ref{eq_dp}) is $\infty$. On the contrary, if $\lc[i-1,\calP,S] < \infty$, then there exists a $(i-1,\calP,S)$-compatible set $C$ which, by Claim~\ref{claim_gamma(iPS)_finite}(1), is a $(i,\calP,S)$-compatible set. This contradicts our assumption in this case and therefore, $\lc[i-1,\calP,S] = \infty$. Similarly, if we assume that $\lc[i-1,\calP',S'] < \infty$ for some refinement $\calP'$ of $\calP_0$ and some subset $S'$ of $T$ such that $\calP' \Cap \calP(r_i) = \calP$ and $S' \cup N(r_i) = S$, then there exists a $(i-1,\calP',S')$-compatible set $C'$. Therefore, by Claim~\ref{claim_gamma(iPS)_finite}(2), the set $C = C' \cup \{r_i\}$ a $(i,\calP,S)$-compatible set. This again implies a contradiction to our assumption in this case and therefore, $\lc[i-1,\calP',S'] = \infty$. Hence, this proves that RHS $= \infty$ in Equation~(\ref{eq_dp}). 

In view of the above argument, therefore, we assume for the 
rest of the proof that $\lc[i,\calP,S] < \infty$. 
We then show that LHS $=$ RHS in 
Equation~(\ref{eq_dp}). To begin with, we first show that LHS $\leq$ RHS in 
Equation~(\ref{eq_dp}). This is true since, by 
Claim~\ref{claim_gamma(iPS)_finite}(1), 
we have $\lc[i,\calP,S] \leq \lc[i-1,\calP,S]$ and, by 
Claim~\ref{claim_gamma(iPS)_finite}(2), 
for any $(i-1,\calP',S') \in \mathcal{X}$ such that 
$\calP' \cap \calP(r_i) = \calP$ and 
$S' \cup N(r_i) = S$, we have $\lc[i,\calP,S] \leq 1+\lc[i-1,\calP',S']$.

We now prove the reverse, that is, the inequality LHS $\geq$ RHS in Equation~(\ref{eq_dp}). 
Recall that $\lc[i,\calP,S] < \infty$ and hence, $\lc[i,\calP,S] = |C|$ for some minimum $(i,\calP,S)$-compatible set $C$. We consider the following two cases.
\begin{itemize}[nolistsep]
\item \textbf{Case 1} ($r_i \notin C$).
In this case, $C$ is a $(i-1,\calP,S)$-compatible set 
which implies that RHS $\leq \lc[i-1,\calP,S] \leq |C| = \lc[i,\calP,S]$.

\item \textbf{Case 2} ($r_i \in C$).
Let $C' = C \setminus \{r_i\}$. 
Then, $C'$ is a $(i-1,\calP',S')$-compatible set, 
where $\calP' \cap \calP(r_i) = \calP$ and 
$S' \cup N(r_i) = S$. 
Therefore, we have $\lc[i-1,\calP',S'] \leq |C'|$ 
which implies that RHS 
$\leq 1+\lc[i-1,\calP',S'] \leq 1+|C'| = |C| = \lc[i,\calP,S]$.
\end{itemize}

This proves that LHS = RHS in Equation~(\ref{eq_dp})
and hence the correctness of the algorithm.

\subparagraph*{Running Time.}
As mentioned before, given any $i \in [\ell]$ and any $S \subseteq T$, 
the number of partitions $\calP$ for which $(i,\calP,S) \in \x$ is 
$2^{\calO(|B| \log |B|)} \cdot |R|$. 
Moreover, the number of possible subsets $S$ of $T$ 
is at most $2^{|T|} \in 2^{\calO(|B|)}$. 
Therefore, there are at most $|R| \cdot 2^{\calO(|B| \log |B|)} 
\cdot |R| \cdot 2^{\calO(|B|)} \in 2^{\calO(|B| \log |B|)} \cdot |R|^{\calO(1)}$
many states of the dynamic programming that need to be calculated.
Moreover, to calculate each such $\lc[i,\calP,S]$, another 
$2^{\calO(|B| \log |B|)}\cdot |R|^{\calO(1)}$ number of 
$\lc[i-1,\calP',S]$'s are  invoked according to the dynamic programming formula~(\ref{eq_dp}).
As a result therefore, we obtain a total runtime of 
$2^{\calO(|B| \log |B|)} \cdot |R|^{\calO(1)} \times 2^{\calO(|B| \log |B|)} \cdot |R|^{\calO(1)}$.
This completes the proof of Lemma~\ref{lemma:partition-refinement-fpt}.

\subsection*{Parameterization by the number of items for \TCPB}

We find it convenient to with the auxiliary graph representation
of the instance $(U,\calF, k)$ of \TCPB.
Recall that we define the {auxiliary bipartite graph} as
graph $G$ on $n$ vertices with bipartition $\langle R, B \rangle$
of $V(G)$ such that sets $R$ and $B$ contain a vertex
for every set in $\calF$ and for every item in $U$, respectively,
and $r \in R$ and $b \in B$ are adjacent 
if and only if the set corresponding to $r$ contains 
the item corresponding to $b$.
\TCPB\ admits a simple $|R|^{\calO(k)} \cdot (|R|
+|B|)^{\calO(1)}$ algorithm that enumerates all the possible subsets
of $R$ and checks it for feasibility. Our goal is find an
\FPT\ algorithm parameterized by $|B|$. 
Note that it is safe to assume
$|R| \le 2^{|B|}$ as we can remove twin vertices from $R$ and get
an equivalent instance).
Bondy's theorem~\cite{bondy1972induced}
asserts that \TCPB always has a solution of size at most
$|B|-1$.
This implies that $k \le |B|$ for a non-trivial instance.
Hence, the brute-force algorithm mentioned above runs in
time $2^{\calO(|B|^2)} \cdot (|R| +|B|)^{\calO(1)}$.

Using Lemma~\ref{lemma:partition-refinement-fpt}, we present an
improved \FPT\ algorithm, in the spirit of the known $\calO(mn2^n)$
dynamic programming scheme for \textsc{Set Cover} on a universe of
size $n$ and $m$ sets (see~\cite[Theorem~3.10]{DBLP:series/txtcs/FominK10}). 
We reduce \TCPB\ to \textsc{Annotated Red-Blue Partition Refinement}. 
For this, note that in a solution of \TCPB, either 
every vertex of $B$ is dominated by the solution, or 
exactly one vertex of $B$ is not dominated by the solution. 
Thus, we go through every possible choice of subset $T_L
\subseteq B$ with $|T_L|\in\{|B|-1,|B|\}$. 
We let $C_0=\emptyset$,
and define the partition $\q$ defined by taking all vertices of $R$ as
parts of size~1, and the set $B$ as the last part. Now, it suffices to
run the dynamic programming for solving \textsc{Annotated Red-Blue
  Partition Refinement} with each of the $|B|+1$ choices of set
$T_L$, giving a running time of $2^{\calO(|B|\log|B|)} \cdot (|R| +
|B|)^{\calO(1)}$.

%% file: other-structural-para.tex
\subsection*{Parameterization by Other Structural Parameters}

We now consider further structural parameterizations of \LD.

\paragraph*{Extending Theorem~\ref{thm:LD-struct-vc} for Twin-Cover Number and Distance to Clique}

The technique designed in Section~\ref{sec:vc-fpt} 
can in fact be used to obtain algorithms for two related parameters. 
The \emph{distance to clique} of a graph is the size of a 
smallest set of vertices that need to be removed from a graph
so that the remaining vertices form a
clique. It is a dense analogue of the vertex cover number (which can
be seen as the ``distance to independent set''). The \emph{twin-cover
number} of a graph~\cite{DBLP:conf/iwpec/Ganian11} is the size of a
smallest set of vertices such that the remaining vertices is a
collection of disjoint cliques, each of them forming a set of mutual
twins. The twin-cover number is at most the vertex cover number, since
the vertices not in the vertex cover form cliques of size~1 in the
remaining graph.
We can obtain running in time $2^{\calO(\dclique\log\dclique)} \cdot n^{\calO(1)}$ and $2^{\calO(\tc\log\tc)} \cdot n^{\calO(1)}$
, where \dclique\ is the distance to clique of the input graph, and \tc\ is
the twin-cover cover number of the input graph. In both cases, the arguments are essentially the
same as for $\vc$.

For the twin-cover parameterization, let $U$ be a set of vertices of
size $\tc$ whose removal leaves a collection $\mathcal C$ of cliques,
each forming a twin-class. Observe that, since we have applied
Reduction Rule~\ref{reduc:twins} exhaustively, in fact every clique in
$\mathcal C$ has size at most~2. Moreover, recall that for any two
vertices in a twin-class $S$, we need to select any one of them in any
solution, thus those vertices are put in $C_0$. Note that each such
selected vertex dominates its twin in $R$, which thus is the only
vertex of $R$ in its part of $\mathcal Q$. Apart from this, nothing
changes, and we can apply Lemma~\ref{lemma:partition-refinement-fpt}
as above.

For distance to clique, $U$ is the set of vertices of size $\dclique$
whose removal leaves a clique, $R$. Since $R$ forms a clique, we do no
longer immediately obtain a bipartite graph, and thus we need a
``split graph'' version of Lemma~\ref{lemma:partition-refinement-fpt},
where $R$ is a clique and $B$ remains an independent set. Here, there
are some slight changes during the phase of pre-processing the
partition in the proof of Lemma~\ref{lemma:partition-refinement-fpt},
since any vertex in $C$ from $R$ dominates the whole of $R$ and can
thus potentially locate some vertex of $R$ from some vertex of
$B$. Nevertheless, after the pre-processing, as in
Lemma~\ref{lemma:partition-refinement-fpt}, we obtain a partition
where every part is completely included either in $R$ or in $B$. Then,
selecting a vertex of $R$ does not distinguish any two other vertices
of $R$, and hence the algorithm is exactly the same from that point
onwards.

\paragraph*{A linear kernel for Neighbourhood Diversity}

We now show that \LD admits a linear kernel when parameterized by the
neighbourhood diversity of the input graph. Recall that the
\emph{neighbourhood diversity} of a graph $G$ is the smallest integer
$d$ such that $G$ can be partitioned into $d$ sets of mutual twin
vertices, each set being either a clique or an independent set. It was
introduced in~\cite{DBLP:journals/algorithmica/Lampis12} and can be
computed in linear time $O(n+m)$ using modular
decomposition~\cite{DBLP:journals/talg/CoudertDP19}.
We show the following.

\begin{lemma}\label{thm:ND-kernel}
\LD admits a kernelization algorithm that, given an instance $(G,k)$ with $nd(G)=d$, computes an equivalent instance $(G',k')$ with $2d$ vertices, and that runs in time $\calO(nm)$.
\end{lemma}
\begin{proof}
We repeatedely apply Reduction Rule~\ref{reduc:twins} until there are no sets of mutual twins of size more than~2 in the graph. As each application takes time $\calO(n+m)$ (Lemma~\ref{lem:reduc-twins}), overall this take at most time $\calO(n(n+m))=\calO(nm)$. Thus, after applying the rule, there are at most two vertices in each neighbourhood equivalence class. Since there are at most $d$ such classes, the resulting graph has at most $2d$ vertices. By Lemma~\ref{lem:reduc-twins}, this produces an equivalent instance $(G',k')$.
\end{proof}

\begin{corollary}\label{cor:ND-FPT}
\LD admits an algorithm running in time $2^{\calO(d)}+\calO(nm)$, where $d$ is the neighbourhood diversity of the input graph.
\end{corollary}
\begin{claimproof}
We first compute in time $\calO(nm)$ the linear kernel from Theorem~\ref{thm:ND-kernel}, that is a graph with $n'\leq 2d$ vertices. Then, a simple brute-force algorithm running in time $2^{\calO(n')}=2^{\calO(d)}$ completes the proof.
\end{claimproof}

%% file: fes-kernel.tex
\section{Parameterization by Feedback Edge Set Number}
\label{sec:fes-kernel}

We next prove Theorem~\ref{thm:LD-struct-fes}. A set of edges $X$ of a graph is a {\em feedback edge set} if $G - X$ is a forest.
The minimum size of a feedback edge set is known as the {\em feedback edge set number} ${\fes}$. 
If a graph $G$ has $n$ vertices, $m$ edges, and $r$ components, then ${\fes}(G) = m - n + r$ \cite{Diestel12}.
%


Our proof
is nonconstructive, in the sense that it shows the \emph{existence} of a large (but constant) number of gadget types with which the kernel can be constructed, however, as there are far too many gadgets, we cannot describe them individually. Missing parts of the proof are found in Appendix~E.
We first describe graphs of given feedback edge set number. 

\begin{proposition}[{\cite[Observation 8]{KK20}}]\label{prop:FES-structure}
Any graph $G$ with feedback edge set number $\fes$ is obtained from a multigraph $\widetilde{G}$ with at most $2\fes-2$ vertices and $3\fes-3$ edges by first subdividing edges of $\widetilde{G}$ an arbitrary number of times, and then, repeatedly attaching degree~1 vertices to the graph, and $\widetilde{G}$ can be computed from $G$ in $\calO(n+\fes)$ time.
\end{proposition}

By Proposition~\ref{prop:FES-structure}, we compute the multigraph $\widetilde{G}$ in $\calO(n+ \fes)$ time, and we let $\widetilde{V}$ be the set of vertices of $\widetilde{G}$. For every edge $v_1v_2$ of $\widetilde{G}$, we have either:
\begin{enumerate}
\item $v_1v_2$ corresponds to an edge in the original graph $G$, if $v_1v_2$ has not been subdivided when obtaining $G$ from $\widetilde{G}$ (then $v_1\neq v_2$ since $G$ is loop-free), or
\item $v_1v_2$ corresponds to a component $C$ of the original graph $G$ with $\{v_1,v_2\}$ removed. Moreover, $C$ induces a tree, with two (not necessarily distinct) vertices $c_1,c_2$ in $C$, where $c_1$ is the only vertex of $C$ adjacent to $v_1$ and $c_2$ is the only vertex of $C$ adjacent to $v_2$ in $G$. (Possibly, $v_1=v_2$ and the edge $v_1v_2$ is a loop of $\widetilde{G}$.)
\end{enumerate}

To avoid dealing with the case of loops, for every component $C$ of $G\setminus\{v_1\}$ with a loop at $v_1$ in $\widetilde{G}$ and with $c_1,c_2$ defined as above, we select an arbitrary vertex $x$ of the unique path between $c_1$ and $c_2$ in $G[C]$ and add $x$ to $\widetilde{V}$. As there are at most $\fes$ loops in $\widetilde{G}$, we can assume that $\widetilde{G}$ is a loopless multigraph with at most $3\fes-2$ vertices and $4\fes-3$ edges.

\begin{observation}\label{obs:noloops}
We may assume that the multigraph $\widetilde{G}$ computed in Proposition~\ref{prop:FES-structure} has at most $3\fes-2$ vertices and $4\fes-3$ edges, and no loops.
\end{observation}
 

The kernelization has two steps. First, we need to handle \emph{hanging trees}, that is, parts of the graph that induce trees and are connected to the rest of the graph via a single edge. Those correspond to the iterated addition of degree 1 vertices to the graph from $\widetilde{G}$ described in Proposition~\ref{prop:FES-structure}. Step~2 will deal with the subgraphs that correspond to the edges of $\widetilde{G}$.

\subparagraph*{Step 1 -- Handling the Hanging Trees.} It is known that there is a linear-time dynamic programming algorithm to solve {\LD} on trees~\cite{SlaterTrees} (see also~\cite{ABLW20} for a more general algorithm for block graphs, that also solves several types of problems related to \LD).
To this end, one can in fact define the following five types of (partial) solutions. For a tree $T$ rooted at a vertex $v$, we define five types for a subset $L$ of $V(T)$ as follows:

\begin{itemize}[nolistsep]
\item Type $A$: $L$ dominates all vertices of $V(T)\setminus \{v\}$ and any two vertices of $V(T)\setminus (L\cup\{v\})$ are located.
\item Type $B$: $L$ is a dominating set of $T$ and any two vertices of $V(T)\setminus (L\cup\{v\})$ are located;
\item Type $C$: $L$ is a locating-dominating set of $T$;
\item Type $D$: $L$ is a locating-dominating set of $T$ with $v\in L$;
\item Type $E$: $L$ is a locating-dominating set of $T$ such that $v\in L$ and there is no vertex $w$ of $T$ with $N(w)\cap L=\{v\}$; 
\end{itemize}

Next, for convenience, we consider the lexicographic ordering $A<B<C<D<E$. Note that these five types of solution are increasingly constrained (so, if $X<X'$, if $L$ is a solution of type $X'$, it is also one of type $X$). 
For a tree $T$ rooted at $v$ and $X\in\{A,B,C,D,E\}$, we denote by $opt_X(T,v)$ the minimum size of a set $L\subseteq V(T)$ of type $X$. Note that $opt_C(T,v)$ is equal to the location-domination number of $T$.

The following proposition can be proved by using the same ideas as in the dynamic programming algorithms from \cite{ABLW20,SlaterTrees}.

\begin{proposition}
\label{prop:DP-trees}
There is a linear-time dynamic programming algorithm that, for any tree $T$ and root $v$, conjointly computes $opt_X(T,v)$ for all $X\in\{A,B,C,D,E\}$.
\end{proposition}




\begin{lemma}
\label{lemma:types-bounded}
  Let $T$ be a tree with vertex $v$, and let $X,X'\in\{A,B,C,D,E\}$ with $X<X'$. We have $opt_{X'}(T,v)\leq opt_{X}(T,v)\leq opt_{X'}(T,v)+1$.
\end{lemma}
\begin{proof}
  The first inequality follows immediately by noticing that the conditions for types $A$ to $E$ are increasingly constrained. Moreover, all types are feasible by considering $L=V(T)$, so $opt_X(T,v)$ is always well-defined. This implies that the first inequality is true.

  For the second inequality, assume we have $X'\in\{A,B\}$ and $L$ is an optimal solution of type $X'$. Then, $L\cup\{v\}$ is a solution of type $E$. If $X'\in\{C,D\}$, let $L$ be an optimal solution of type $X'$. If $v\notin L$ (then $X'=C$), then again $L\cup\{v\}$ is a solution of type $E$. If $v\in L$, possibly some (unique) vertex $w$ of $T$ satisfies $N(w)\cap L=\{v\}$. Then, $L\cup\{w\}$ is a solution of type $E$. Thus, $opt_{X}(T,v)\leq opt_E(T,v)\leq opt_{X'}(T,v)+1$. In all cases, we have $opt_{X}(T,v)\leq opt_E(T,v)\leq opt_{X'}(T,v)+1$, as claimed. 
\end{proof}

By Lemma~\ref{lemma:types-bounded}, for any rooted tree $(T,v)$, the
five values $opt_{X}(T,v)$ (for $X\in\{A,B,C,D,E\}$) differ by at most
one, and we know that they increase along with $X$: $opt_{A}(T,v)\leq opt_{B}(T,v)\leq \ldots \leq opt_{E}(T,v)$. Hence, for any rooted tree $(T,v)$, there is one value
$X\in\{A,B,C,D,E\}$ 
so that, whenever $X'\leq X$, we have $opt_{X'}(T,v)=opt_{X}(T,v)$, and
whenever $X'>X$, we have $opt_{X'}(T,v)=opt_{X}(T,v)+1$. Thus, we can
introduce the following definition, which partitions the set of rooted
trees according to one of five such possible behaviours.

\begin{definition}[Rooted tree classes]
For $X\in\{A,B,C,D,E\}$, we define the class $\mathcal T_X$ as the set
of pairs $(T,v)$ such that $T$ is a tree with root $v$, and there is
an integer $k$ with $opt_{X'}(T,v)=k$ whenever $X'\leq X$, and
$opt_{X'}(T,v)=k+1$ whenever $X'>X$ (note that $k$ may differ across
the trees in $\mathcal T_X$).
\end{definition}

Note that for the trees in $\mathcal T_E$, all five types of optimal
solutions have the same size.
  
It is clear by Lemma~\ref{lemma:types-bounded} that the set of all possible pairs $(T,v)$ where $T$ is a tree and $v$ a vertex of $T$, can be partitioned into the five classes $\mathcal T_X$, where $X\in\{A,B,C,D,E\}$. Moreover, it is not difficult to construct small rooted trees for each of these five classes. We provide five such small rooted trees in Table~\ref{tab:tree-gadgets}, each in a different class. 

\def\myfigscale{0.8}

\begin{table} \centering  
  \begin{tabular}{|c|c|c|c|c|c|}

 \hline

    &

    \scalebox{\myfigscale}{\begin{tikzpicture}[join=bevel,inner sep=0.5mm]
        \node (v) at (0,0) [draw, circle, label={0:$v_A$}] {}; 
        \node (e) at (0,-1.5) {};
    \end{tikzpicture}}

    &

    \scalebox{\myfigscale}{\begin{tikzpicture}[join=bevel,inner sep=0.5mm]
        \node (v) at (0,0) [draw, circle, label={0:$v_B$}] {}; 
        \node (a) at (0,-.5) [draw, circle] {};
        \node (b) at (0,-1) [draw, circle] {};
        \node (e) at (0,-1.5) {};
        \draw [-] (b)--(a)--(v);
    \end{tikzpicture}}
    
    &

    \scalebox{\myfigscale}{\begin{tikzpicture}[join=bevel,inner sep=0.5mm]
        \node (v) at (0,0) [draw, circle, label={0:$v_C$}] {}; 
        \node (a) at (-.25,-.5) [draw, circle] {};
        \node (b) at (-.5,-1) [draw, circle] {};
        \node (c) at (.25,-.5) [draw, circle] {};
        \node (d) at (.5,-1) [draw, circle] {};
        \node (e) at (0,-1.5) {};
        \draw [-] (b)--(a)--(v)--(c)--(d);
    \end{tikzpicture}}

    &
    \scalebox{\myfigscale}{\begin{tikzpicture}[join=bevel,inner sep=0.5mm]
        \node (v) at (0,0) [draw, circle, label={0:$v_D$}] {}; 
        \node (a) at (0,-.5) [draw, circle] {};
        \node (e) at (0,-1.5) {};
        \draw [-] (a)--(v);
    \end{tikzpicture}}

    &
    
    \scalebox{\myfigscale}{\begin{tikzpicture}[join=bevel,inner sep=0.5mm]
        \node (v) at (0,0) [draw, circle, label={0:$v_E$}] {}; 
        \node (a) at (-.25,-.5) [draw, circle] {};
        \node (b) at (-.5,-1) [draw, circle] {};
        \node (c) at (.25,-.5) [draw, circle] {};
        \node (d) at (.5,-1) [draw, circle] {};
        \node (e) at (.75,-1.5) [draw, circle] {};
        \draw [-] (b)--(a)--(v)--(c)--(d)--(e);
    \end{tikzpicture}}\\

  & $T_A$ & $T_B$  & $T_C$ & $T_D$ & $T_E$\\

\hline
$opt_A$ & 0 & 1 & 2 & 1& 3\\
\hline
$opt_B$ & 1 & 1 & 2 & 1 & 3\\
\hline
$opt_C$ & 1 & 2 & 2 & 1 & 3\\
\hline
$opt_D$ & 1 & 2 & 3 & 1 & 3\\
\hline
$opt_E$ & 1 & 2 & 3 & 2 & 3\\
\hline
  \end{tabular}
  \caption{The five rooted tree gadgets with their values of the five $opt_X$ functions.}
  \label{tab:tree-gadgets}
\end{table}

\begin{definition}[Rooted tree gadgets]\label{def:tree-gadgets}
For $X\in\{A,B,C,D,E\}$, let $(T_X,v_X)$ be the rooted tree of class $\mathcal T_X$ from Table~\ref{tab:tree-gadgets}, and let $k_X=opt_X(T_X,v_X)$.
\end{definition}




We are now ready to present our first reduction rule.

\begin{reduction rule}
\label{reduc:hanging-tree}
Let $(G,k)$ be an instance of \LD such that $G$ can be obtained from a graph $G'$ and a tree $T$, by identifying a vertex $v$ of $T$ with a vertex $w$ of $G'$, such that $(T,v)$ is in class $\mathcal T_X$ with $X\in\{A,B,C,D,E\}$, and with $opt_X(T,v)=t$.
Then, we remove all vertices of $T\setminus\{v\}$ from $G$ (this results in $G'$), we consider $G$ and a copy of the rooted tree gadget $(T_X,v_X)$, and we identify $v_X$ with $v$, to obtain $G''$.
The reduced instance is $(G'',k-t+k_X)$.
\end{reduction rule}

\begin{lemma}\label{lemm:reduc-hanging-tree}
For an instance $(G,k)$ of \LD, Reduction Rule~\ref{reduc:hanging-tree} can be applied exhaustively in time $\calO(n^2)$, and for each application, $(G,k)$ is a YES-instance of \LD if and only if $(G'',k-t+k_X)$ is a YES-instance of \LD.
\end{lemma}

\begin{proof}
  To apply Reduction Rule~\ref{reduc:hanging-tree}, one can find all vertices of degree~1 in $G$, mark them, and iteratively mark degree~1 vertices in the subgraph induced by unmarked vertices. Note that a connected subset $S$ of marked vertices induces a tree with a single vertex $v_S$ adjacent to a non-marked vertex $w$. Moreover, at the end of the marking process, the subgraph induced by non-marked vertices has minimum degree~2. This process can be done in $\calO(n^2)$ time.

  We form the set of candidate trees $T$ by taking the union $S_w$ of all marked subsets that are connected to some common unmarked vertex $w$, together with $w$. Note that all these candidate trees are vertex-disjoint. Again, this can be done in $\calO(n^2)$ time. Let $(T,v)$ be such a rooted tree, and let $G'$ be the subgraph induced by $V(G)\setminus S_v$. Then, we apply the linear-time dynamic programming algorithm of Proposition~\ref{prop:DP-trees} to $(T,v)$ to determine the class $\mathcal T_X$ to which $(T,v)$ belongs to, for some $X\in\{A,B,C,D,E\}$, as well as $opt_X(T,v)=t$ for each $X\in\{A,B,C,D,E\}$. Constructing $(G'',k-t+k_X)$ by replacing $T$ by $T_X$ in $G$ can then be done in time $\calO(|V(T)|)$. Thus, overall applying the reduction rule to all trees takes $\calO(n^2)$ time.

  Now, we prove the second part of the statement. Assume we have applied Reduction Rule~\ref{reduc:hanging-tree} to a single tree $(T,v)$ to $G$,that $(G,k)$ is a YES-instance of \LD, and let $L$ be an optimal locating-dominating set of $G$ of size at most $k$. Assume that $L\cap V(T)$ is of type $X'$ for $X'\in\{A,B,C,D,E\}$ with respect to $(T,v)$. Notice that the only interactions between the vertices in $V(T)$ and $V(G')$ are through the cut-vertex $v$. If $X'\leq X$, then, as $L$ is optimal, we have $|L\cap V(T)|=t$; if $X'>X$, then by Lemma~\ref{lemma:types-bounded}, we have $|L\cap V(T)|=t+1$. Now, we obtain a solution of $G''$ from $L'=L\cap V(G')$ by adding an optimal solution $S_{X'}$ of type $X'$ of $(T_X,v_X)$ to $L'$. This has size $k_X$ if $X'\leq X$ and $k_X+1$ if $X'>X$. As $S_{X'}$ behaves the same as $L\cap V(T)$ with respect to $L'$, $L$ is a valid locating-dominating set of $G''$. Its size is either $|L|-(t+1)+k_X+1$ if $X'>X$, or $|L|-t+k_X$ if $X\leq X'$, which in both cases is at most $k-t+k_X$, thus $(G'',k-t+k_X)$ is a YES-instance.

  The proof of the converse follows by the same argument: essentially, the trees $(T,v)$ and $(T_X,v_X)$ are interchangeable.
\end{proof}

We now use Lemma~\ref{lemm:reduc-hanging-tree} to apply Reduction Rule~\ref{reduc:hanging-tree} exhaustively in time $\calO(n^2)$, before proceeding to the second part of the algorithm.

\subparagraph*{Step 2 -- Handling the subgraphs corresponding to edges of $\widetilde{G}$.} In the second step of our algorithm, we construct tree gadgets similar to those defined in Step~1, but with \emph{two} distinguished vertices. We thus extend the above terminology to this setting. Let $(T,v_1,v_2)$, be a tree $T$ with two distinguished vertices $v_1,v_2$. We define types of trees as in the previous subsection. To do so, we first define the types for each of $v_1$ and $v_2$, and then we combine them. For $X\in\{A,B,C,D,E\}$, a subset $L$ of $V(T)$ is of \emph{type $(X,-)$ with respect to $v_1$} if: 
\begin{itemize}[nolistsep]
\item Type $(A,-)$: $L$ dominates all vertices of $V(T)\setminus \{v_1,v_2\}$ and any two vertices of $V(T)\setminus (L\cup\{v_1,v_2\})$ are located.
\item Type $(B,-)$: $L$ dominates all vertices of $V(T)\setminus \{v_2\}$ and any two vertices of $V(T)\setminus (L\cup\{v_1,v_2\})$ are located;
\item Type $(C,-)$: $L$ dominates all vertices of $V(T)\setminus \{v_2\}$ and any two vertices of $V(T)\setminus (L\cup\{v_2\})$ are located;
\item Type $(D,-)$: $L$ dominates all vertices of $V(T)\setminus \{v_2\}$, $v_1\in L$, and any two vertices of $V(T)\setminus (L\cup\{v_2\})$ are located;
\item Type $(E,-)$: $L$ dominates all vertices of $V(T)\setminus \{v_2\}$, $v_1\in L$, any two vertices of $V(T)\setminus (L\cup\{v_2\})$ are located, and there is no vertex $w$ of $V(T)\setminus\{v_2\}$ with $N(w)\cap L=\{v_1\}$;
\end{itemize}

Thus, the idea for a set $L$ of type $(X,-)$ with respect to $v_1$ is that there are specific constraints for $v_1$, but no constraint for $v_2$ (and the usual domination and location constraints for the other vertices). 
We say that $L$ if of type $(X,Y)$ with respect to $(T,v_1,v_2)$ if $L$ is of type $(X,-)$ with respect to $v_1$ and of type $(Y,-)$ with respect to $v_2$. Thus, we have a total of 25 such possible types. 

We let $opt_{X,Y}(T,v_1,v_2)$ be the smallest size of a subset $L$ of $T$ of type $(X,Y)$. 
Here as well, there is a linear-time dynamic programming algorithm that, for any tree $T$ and two vertices $v_1,v_2$, conjointly computes $opt_{X,Y}(T,v_1,v_2)$ for all $X,Y\in\{A,B,C,D,E\}$.

\begin{proposition}\label{prop:DP-trees-2}
There is a linear-time dynamic programming algorithm that, for any tree $T$ and two vertices $v_1,v_2$, conjointly computes $opt_{X,Y}(T,v_1,v_2)$ for all $X,Y\in\{A,B,C,D,E\}$.
\end{proposition}

\begin{proof}
The dynamic programming is similar to that of Proposition~\ref{prop:DP-trees}. We root $T$ at $v_1$. The main difference is that extra care must be taken around $v_2$ to only keep the solutions of type $(-,Y)$ for the subtrees that contain $v_2$.
\end{proof}

Next, similarly to Lemma~\ref{lemma:types-bounded}, we show that for a triple $(T,v_1,v_2)$, the optimal solution values for different types, may vary by at most~2.

\begin{lemma}\label{lemma:types-bounded-2}
  Let $T$ be a tree with vertices $v_1,v_2$ of $T$, and let $X,X',Y,Y'\in\{A,B,C,D,E\}$ with $X\leq X'$ and $Y\leq Y'$. If $X=X'$ and $Y<Y'$ or $X<X'$ and $Y=Y'$, we have $opt_{X,Y}(T,v_1,v_2)\leq opt_{X',Y'}(T,v_1,v_2)\leq opt_{X,Y}(T,v_1,v_2)+1$. If $X<X'$ and $Y<Y'$, we have $opt_{X,Y}(T,v_1,v_2)\leq opt_{X',Y'}(T,v_1,v_2)\leq opt_{X,Y}(T,v_1,v_2)+2$.
\end{lemma}
\begin{proof}
  The proof is similar to that of Lemma~\ref{lemma:types-bounded}. As before, the lower bound is clear since the conditions for type $(X',Y')$ are more constrained that those for $(X,Y)$. For the upper bound, assume first that $X=X'$ or $Y=Y'$ (without loss of generality, $Y=Y'$). Then, as before, given a solution $S$ of type $(X,Y)$, we can add to $S$ a single vertex in the closed neighbourhood of $v_1$ to obtain a solution of type $(E,Y)$. Similarly, if both $X<X'$ and $Y<Y'$, we can add to $S$ a vertex to the closed neighbourhoods of each of $v_1$ and $v_2$ and obtain a solution of type $(E,E)$.
\end{proof}

By Lemma~\ref{lemma:types-bounded-2}, we can introduce the following definition. Note that for a triple $(T,v_1,v_2)$, the minimum value of $opt_{X,Y}(T,v_1,v_2)$ over all $X,Y\in\{A,B,C,D,E\}$ is reached for $X=Y=A$ (and the maximum, for $X=Y=E$).

\begin{definition}[Doubly rooted tree classes]
  For a function $g:\{A,B,C,D,E\}^2\to\{0,1,2\}$, we define the class $\mathcal T_{g}$ as the set of triples $(T,v_1,v_2)$ such that $T$ is a tree, $v_1,v_2$ are two of its vertices, and for every $(X,Y)\in\mathcal\{A,B,C,D,E\}^2$, we have $opt_{X,Y}(T,v_1,v_2)-opt_{A,A}(T,v_1,v_2)=g(X,Y)$.
\end{definition}

There are at most $3^{25}$ possible classes of triples, since that is the number of possible functions $g$. By Lemma~\ref{lemma:types-bounded-2}, these classes define a partition of all sets of triples $(T,v_1,v_2)$.\footnote{Some of these classes are actually empty, since, for example, by Lemma~\ref{lemma:types-bounded-2}, $opt_{X,Y}(T,v_1,v_2)$ increases with $X$ and $Y$; however, an empirical study shows that the number of nonempty classes is very large.}

We next define suitable tree gadgets for the above classes of triples. As there is a very large number of possible types of triples, we cannot give a concrete definition of such a gadget, but rather, an existential one.

\begin{definition}[Doubly-rooted tree gadgets]\label{def:double-gadget}
  For a function $g:\{A,B,C,D,E\}^2\to\{0,1,2\}$, let $(T_{g},v_X,v_Y)$ be the \emph{smallest} tree in $\mathcal T_{g}$ (if such a tree exists), and let $k_{g}=opt_{A,A}(T_{g},v_X,v_Y)$.
\end{definition}

We are now ready to present our second reduction rule, which replaces certain triples $(T,v_1,v_2)$ by gadgets from Definition~\ref{def:double-gadget}.

\begin{reduction rule}\label{reduc:hanging-tree-2}
  Let $(G,k)$ be an instance of \LD, with:
  \begin{itemize}
  \item an induced subgraph $G'$ of $G$ with two vertices $w_1,w_2$;
  \item a tree $T$ with two vertices $v_1,v_2$ with $(T,v_1,v_2)$ in class $\mathcal T_{g}$ with $g:\{A,B,C,D,E\}^2\to\{0,1,2\}$, and with $opt_{A,A}(T,v_1,v_2)=t$;
  \item $w_1,w_2$ do not belong to a hanging tree of $G$;
  \item $G$ can be obtained by taking a copy of $T$ and a copy of $G'$ and identifying, respectively, $v_1$ with $w_1$, and $v_2$ with $w_2$.
  \end{itemize}

Then, we remove all vertices of $T\setminus\{v_1,v_2\}$ from $G$ (that is, we compute $G'$), we consider $G'$ and a copy of doubly-rooted tree gadget $(T_{g},v_X,v_Y)$, and we identify $v_X$ with $v_1$ and $v_Y$ with $v_2$, to obtain $G''$. The reduced instance is $(G'',k-t+k_{g})$.
\end{reduction rule}

\begin{lemma}\label{lemm:reduc-hanging-tree-2}
For an instance $(G,k)$ of \LD, Reduction Rule~\ref{reduc:hanging-tree-2} can be applied to every tree component corresponding to an edge of $\widetilde{G}$ in time $\calO(n\fes)$, and $(G,k)$ is a YES-instance of \LD if and only if $(G'',k-t+k_{g})$ is a YES-instance of \LD.  
\end{lemma}
\begin{proof}
  To apply the reduction rule as stated, we compute the multigraph $\widetilde{G}$ in time $\calO(n+\fes)$ using Proposition~\ref{prop:FES-structure}; by Observation~\ref{obs:noloops}, we can assume it has no loops and at most $4\fes-3$ edges. Now, for every edge $xy$ of $\widetilde{G}$, we select the neighbour $v_1$ of $x$ and the neighbour $v_2$ of $y$ in $G$, that lie on the $xy$-path corresponding to the edge $xy$ of $\widetilde{G}$. The tree $(T,v_1,v_2)$ is the subgraph of $G$ corresponding to the edge $xy$ of $\widetilde{G}$, and $G'$ is the subgraph of $G$ obtained by removing $T\setminus\{v_1,v_2\}$ from $G$. Computing all these subgraphs can be done in time $\calO(n\fes)$. Computing the type $\mathcal{T}_g$ of $(T,v_1,v_2)$ can be done in time $\calO(|T|)$ by Proposition~\ref{prop:DP-trees-2}, so overall, the running time of this step is $\calO(n\fes)$.

  Note that all the computed subgraphs are vertex-disjoint, we can thus now apply Reduction Rule~\ref{reduc:hanging-tree-2} in time $\calO(|T|)$ to each tree $T$, noting that the gadget $T_g$ is of constant size.
  
For the second part of the statement, the arguments are the same as those of Lemma~\ref{lemm:reduc-hanging-tree}: since they belong to the same class $\mathcal T_{g}$, the trees $(T,v_1,v_2)$ and $(T_{g},v_X,v_Y)$ have the same behaviour with respect to \LD and the overall solution for the graph.
\end{proof}

\subparagraph*{Completion of the proof.} By Lemma~\ref{lemm:reduc-hanging-tree} and Lemma~\ref{lemm:reduc-hanging-tree-2}, we can apply Reduction Rule~\ref{reduc:hanging-tree} and Reduction Rule~\ref{reduc:hanging-tree-2} in time $\calO(n^2+n\fes)=\calO(mn)$. Since $\widetilde{G}$ has at most $3\fes-2$ vertices and $4\fes-3$ edges, we replaced each vertex of $\widetilde{G}$ by a constant-size tree-gadget from Definition~\ref{def:tree-gadgets} and each edge of $\widetilde{G}$ by a constant-size doubly-rooted tree gadget from Definition~\ref{def:double-gadget}, the resulting graph has $\calO(\fes)$ vertices and edges, as claimed. The running time is $2^{\calO(\fes)} + n^{\calO(1)}$ by first computing the kernel and then solving \LD\ in time $2^{\calO(n)}=2^{\calO(\fes)}$ on that kernel. This completes the proof. 

%% file: incompressibility.tex
\section{Incompressiblity of {\LD}}
\label{sec:incompressibility}

In this section, we prove Thereom~\ref{thm:loc-dom-set-incompressibility}
for \LD and {\TCPB}, i.e., we prove that neither \LD\ admits 
a polynomial compression of size $\calO(n^{2 - \epsilon})$ 
for any $\epsilon > 0$, unless $\NP \subseteq \coNP/poly$.
We transfer the incompressibility of 
\textsc{Dominating Set} to \LD.
It will be convenient to reduce
from the \textsc{Red-Blue Dominating Set} problem,
a restricted version of
\textsc{Dominating Set}, which also
does not admit a compression with 
$\calO(n^{2 - \epsilon})$ bits unless 
$\NP \subseteq \coNP/poly$~\cite[Proposition~2]{DBLP:journals/tcs/AgrawalKST21}.
The input to the \textsc{Red-Blue Dominating Set} is a bipartite graph $G'$ with bipartition $\langle R', B'\rangle$ and an integer $k$.
The question asks whether there is a set $S' \subseteq R'$ of at most $k'$ vertices such that $B' \subseteq N(S')$.


It is safe to assume that $G'$ does not contain an isolated vertex.
Indeed, if there is an isolated vertex in $R$, one can delete it 
to obtain an equivalent instance.
If $B$ contains an isolated vertex, it is a trivial \no-instance.
Moreover, it is also safe to assume that $|R'| \le 2^{|B'|}$
as it is safe to remove one of the two false twins in $R'$.
Also, $k' \le |R'|$ as otherwise given instance is a trivial
\yes-instance.

\begin{figure}[t]
\centering
\includegraphics[scale=0.60]{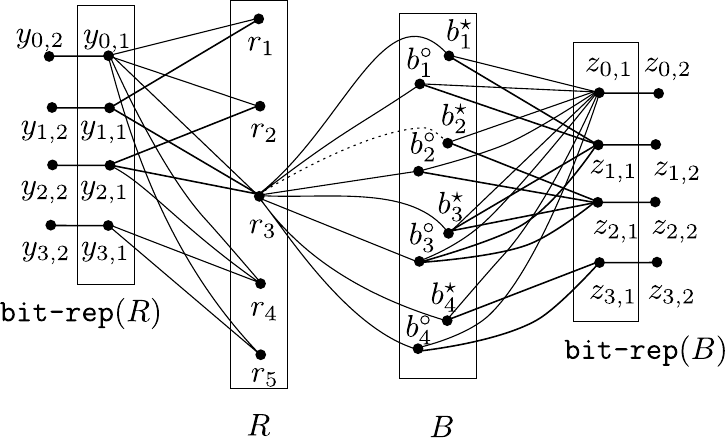}
\caption{An illustrative example of the graph constructed by the reduction used in Section~\ref{sec:incompressibility}.
Adjacency to $y_{1,1}$, $y_{2, 1}$, $y_{3, 1}$ correspond to 
$1$ in the bit representation from left to right (most significant to least significant) bits.
In the above example, the bit representation of
$r_1$ is $\langle 1, 0, 0\rangle$,
$r_2$ is $\langle 0, 1, 0\rangle$, 
$r_3$ is $\langle 1, 1, 0\rangle$, etc.
For brevity, we only show the edges incident on $r_3$.
The dotted line represents a non-edge.
In this example, $r'_3$ is only adjacent to $b'_2$ in $G'$.
It is easy to see that $r_3$ can only resolve pair $(b^{\circ}_2, b^{\star}_2)$.
\label{fig:locating-dom-set-incompressibility}}
\vspace{-3mm}
\end{figure}

\subparagraph{Reduction.}
The reduction takes as input
an instance $(G', \langle R', B' \rangle, k')$ of 
\textsc{Red-Blue Dominating Set} and
constructs an instance $(G, k)$ of 
\LD.
Suppose we have $R' = \{r'_1, r'_2, \dots, r'_{|R|}\}$
and $B' = \{b'_1, b'_2, \dots, b'_{|B|}\}$.
The reduction constructs graph $G$ in the following steps.
See Figure~\ref{fig:locating-dom-set-incompressibility}
for an illustration.
\begin{itemize}[nolistsep]
\item  
It adds sets of vertices $R$ and $B$ to $G$,
where $R$ contains a vertex corresponding 
to each vertex in $R'$ and
$B$ contains two vertices corresponding
to each vertex in $B'$.
Formally, 
$R = \{r_i |\ i \in [|R'|]\}$
and
$B = \{b^{\circ}_i, b^{\star}_i |\ i \in [|B'|]\}$.

\item The reduction adds a \emph{bit representation gadget} to locate set $R$.
Informally, it adds some supplementary vertices such that
it is safe to assume these vertices are present in a locating-dominating set, and they locate every vertex in $R$.
\begin{itemize}[nolistsep]
\item First, set $q := \lceil \log(|R|) \rceil+1$.
This value for $q$ allows to uniquely represent each integer in $[|R|]$ by 
its bit-representation in binary when started from $1$ and not $0$. 
\item For every $i \in [q]$, the reduction adds two vertices $y_{i, 1}$ and $y_{i, 2}$ and edge $(y_{i, 1},  y_{i, 2})$.
\item For every integer $\ell \in [|R|]$, let 
$\bit(\ell)$ denote the binary representation 
of $\ell$ using $q$ bits.
It connects $r_{\ell} \in R$ with $y_{i, 1}$ if the 
$i^{th}$ bit in 
$\bit(\ell)$ is $1$.
\item It adds two vertices $y_{0, 1}$ and $y_{0, 2}$, 
and edge $(y_{0, 1}, y_{0, 2})$, and  
makes every vertex in $R$ adjacent with $y_{0, 1}$.

Let $\bitrep(R)$ be the collection of the vertices {$y_{i,1}$ for all $i \in \{0\} \cup [q]$} added in this step.
\end{itemize}

\item Similarly, the reduction adds a bit representation gadget to locate set $B$.
However, it adds vertices such that for any pair 
$b^{\circ}_j, b^{\star}_j$, the supplementary 
vertices adjacent to them are identical.
\begin{itemize}[nolistsep]
\item The reduction sets $p := \lceil \log(|B|/2) \rceil+1$ and
for every $i \in [p]$, it adds two vertices $z_{i, 1}$ and $z_{i, 2}$ and edge $(z_{i, 1},  z_{i, 2})$.
\item For every integer $j \in [|B|/2]$, let $\bit(j)$ denote the binary representation of $j$ using $p$ bits.
Connect $b^{\circ}_{j}, b^{\star}_j \in B$ with $z_{i, 1}$ if the $i^{th}$ bit in $\bit(j)$ is $1$.
\item It add two vertices $z_{0, 1}$ and $z_{0, 2}$, and edge $(z_{0, 1}, z_{0, 2})$.
It also makes every vertex in $B$ adjacent with $z_{0, 1}$.

Let $\bitrep(B)$ be the collection of the vertices {$z_{i,1}$ for all $i \in \{0\} \cup [p]$} added in this step.
\end{itemize}

\item Finally, the reduction adds edges 
across $R$ and $B$ as follows:
For every pair of vertices $r'_i \in R'$
and $b'_j \in B'$, 
\begin{itemize}[nolistsep]
\item if $r'_i$ is \emph{not} adjacent with $b'_j$ then  it adds both edges
$(r_i, b^{\circ}_j)$ and $(r_i, b^{\star}_j)$ to $E(G)$,
and
\item if $r'_i$ is adjacent with $b'_j$
then it adds $(r_i, b^{\circ}_j)$ only, i.e.,
it does not add an edge with endpoints $r_i$ and 
$b^{\star}_j$.
\end{itemize}
\end{itemize}
This completes the construction of 
$G$.
The reduction sets 
$$k = k' + |\bitrep(R)| + |\bitrep(B)| =  k' + \lceil \log(|R|) \rceil + 1 + 1 + \lceil \log(|B|/2) \rceil + 1 + 1,$$
and returns $(G, k)$ as an instance of 
\LD.

\begin{lemma}
\label{lemma:RBDS-LDS-reduction}
$(G', \langle R', B'\rangle, k')$ is a \yes-instance 
of \textsc{Red-Blue Dominating Set}
if and only if $(G, k)$ is a \yes-instance of
\LD.
\end{lemma}
\begin{proof}
$(\Rightarrow)$
Suppose $(G', \langle R', B'\rangle, k')$
is a \yes-instance and let $S' \subseteq R'$ 
be a solution.
We prove that $S = S' \cup \bitrep(R) \cup \bitrep(B)$ is a locating-dominating set of size at most $k$.
The size bound follows easily. 
By the property of a set representation gadget,
every vertex in $R$ is adjacent with a unique set of vertices 
in $\bitrep(R) \setminus \{y_{0, 1}\}$.
Consider a vertex $r_i$ in $R$ such that the bit-representation
of $i$ contains a single $1$ at $j^{th}$ location.
Hence, $y_{j, 2}$ and $r_i$ are adjacent with the same 
vertex, viz $y_{j, 1}$ in  $\bitrep(R) \setminus \{y_{0, 1}\}$.
However, this pair of vertices is resolved by $y_{0, 1}$ which 
is adjacent with $r_i$ and \emph{not} with $y_{j, 2}$.
Also, as the bit-representation of vertices starts from $1$,
there is no vertex in $R$ which is adjacent with only $y_{0, 1}$ in 
$\bitrep(R)$.
Using similar arguments and the properties of the 
set representation gadget, we can conclude that 
$\bitrep(R) \cup \bitrep(B)$ resolves all pairs of vertices, 
apart from those pairs of the form $(b^{\circ}_j, b^{\star}_j)$. 

By the construction of $G$,
a vertex $r_i \in R$ resolves a pair vertices $b_j^{\circ}$
and $b_j^{\star}$
if and only if $r'_i$ and $b'_j$ are adjacent. 
Since $S'$ dominates $B$,
$S'$ resolves every pair of the form $(b_j^{\circ}, b_j^{\star})$.
Hence, $S' \cup \bitrep(R) \cup \bitrep(B) $ is a 
locating-dominating set of $G$ of the desired size.

$(\Leftarrow)$
Suppose $S$ is a locating-dominating set of $G$ of size at most $k$.
By Observation~\ref{obs:nbr-of-pendant-vertex-in-sol}, 
it is safe to consider that $S$ contains $\bitrep(R) \cup \bitrep(B)$,
as every vertex in it is adjacent to a pendant vertex.

Next, we modify $S$ to obtain another locating-dominating set $S_1$
such that $|S_1| \le |S|$ and
all vertices in $S_1 \setminus (\bitrep(R) \cup \bitrep(B))$
are in $R$.
Without loss of generality, suppose $b^{\circ}_j \in S$.
As discussed in the previous paragraph, 
$\bitrep(R) \cup \bitrep(B)$ resolves all the pair of vertices
apart from the pair of the form $(b_{j'}^{\circ}, b_{j'}^{\star})$.
As there is no edge between the vertices in $B$,
if $b^{\circ}_j \in S$ then it is useful only to resolve the pair
$(b_{j}^{\circ}, b_{j}^{\star})$. 
As there is no isolated vertex in $G'$, every vertex in $B'$
is adjacent with some vertex in $R'$.
Hence, by the construction of $G$,  there is
a vertex $r_i$ in $R$ such that $r_i$ is adjacent
with $b^{\circ}_j$ but not with $b^{\star}_j$.
Consider set $S_1 = (S \setminus \{b_j^{\circ}\}) \cup {r_i}$.
By the arguments above, $|S_1| \le |S|$, $S_1$ is a locating-dominating set of $G$, and $|S_1 \cap B| < |S \cap B|$.
Repeating this step, we obtain set $S_1$ with desired properties.

It is easy to verify that $|S_1 \cap R| = |S_1 \cap R'| \le k'$ and that
for every vertex $b_j$ in $B'$, there is a vertex in $S_1 \cap R'$ which is adjacent with $b_j$.
Hence, the size of $S_1 \cap R$ is at most $k$, and it dominates
every vertex in $B$.
\end{proof}

\begin{proof}[Proof of Theorem~\ref{thm:loc-dom-set-incompressibility}
for \LD]
Suppose there exists a polynomial-time algorithm
$\calA$ that takes as input an instance $(G, k)$ of 
\LD and produces
an equivalent instance $I'$, which requires $\calO(n^{2 - \epsilon})$ bits to 
encode, of some problem $\Pi$.
Consider the compression algorithm $\calB$
for \textsc{Red-Blue Dominating Set} that uses the reduction
mentioned in this section and then the algorithm $\calA$ on
the reduced instance.
For an instance $(G', \langle R', B' \rangle, k')$ of
\textsc{Red-Blue Dominating Set}, where the number of vertices
in $G'$ is $n'$, the reduction constructs a graph $G$ with 
at most $3n'$ vertices.
Hence, for this input algorithm, $\calB$ constructs 
an equivalent instance with $\calO(n'^{2 - \epsilon})$ bits,
contradicting $\NP \subseteq \coNP/poly$~\cite[Proposition~2]{DBLP:journals/tcs/AgrawalKST21}.
Hence, 
\LD does not admit a 
polynomial compression of size
$\calO(n^{2 - \epsilon})$ for any $\epsilon > 0$, 
unless $\NP \subseteq \coNP/poly$.
\end{proof}

\subparagraph{Other Consequences of the Reduction.}
Note that $\bitrep(R) \cup \bitrep(B) \cup B$ is a vertex cover of 
$G$ of size $\log(|R|) + \log(|B|) + |B|$.
As $k' \le |B'|$ (as otherwise we are dealing with a 
trivially \yes-instance of \textsc{Red-Blue Dominating Set}.), 
$|R'| \le 2^{|B'|}$,
we have $\vc(G) + k \le (\log(|R|) + \log(|B|) + |B|) + k' + \calO(\log(|R|) + \log(|B|) + |B|) = \calO(|B|)$. 
It is known that \textsc{Red-Blue Dominating Set},
parameterized by the number of blue vertices, does not admit
a polynomial kernel unless $\NP \subseteq \coNP/poly$.
See, for example, \cite[Lemma~$15.19$]{cygan2015parameterized}.
This, along with the arguments that are standard to
parameter-preserving reductions, implies that \LD,
when parameterized jointly by the solution size and the vertex cover number
of the input graph,
does not admit a polynomial
kernel unless $\NP \subseteq \coNP/poly$.

\subsection*{Incompressibility of Test Cover}
Now, we prove Theorem~\ref{thm:loc-dom-set-incompressibility}
for \TCPB.
Our reduction is a modification (and also a simplification)
of the reduction mentioned in Section~\ref{sec:incompressibility}.
We transfer the incompressibility of 
\textsc{Red-Blue Dominating Set} to \TCPB.
Once again, it is safe to assume that $G'$ does not contain 
an isolated vertex, $R' \le 2^{|B'|}$, and $k' \le |B'|$.

\subparagraph*{Reduction}
For notational convenience, instead of specify an instance
of \TCPB, we specify the auxiliary graph as mentioned 
in the definition.
The reduction takes as input
an instance $(G', \langle R', B' \rangle, k')$ of 
\textsc{Red-Blue Dominating Set} and
constructs an instance $(G, \langle R, B \rangle, k)$ of 
\TCPB.
Suppose we have $R' = \{r'_1, r'_2, \dots, r'_{|R|}\}$
and $B' = \{b'_1, b'_2, \dots, b'_{|B|}\}$.
The reduction constructs graph $G$ in the following steps.
See Figure~\ref{fig:locating-dom-set-incompressibility}
for an illustration.

\begin{figure}[t]
\centering
\includegraphics[scale=0.650]{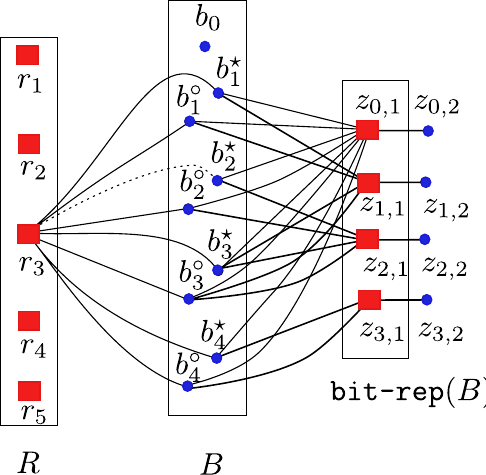}
\caption{An illustrative example of the graph constructed by the reduction for \TCPB.
Red (squared) nodes denote the tests whereas blue (filled circle) nodes
the items.
The adjacencies across $B$ and $\bitrep(B)$ are the same as the ones from the reduction for \LD. For brevity, we only show the edges incident on $r_3$.
The dotted line represents a non-edge.
In this example, test corresponding to $r_3$ only contains item
corresponding to $b_2$.
It is easy to see that $r_3$ can only distinguish between the pair $(b^{\circ}_2, b^{\star}_2)$.
\label{fig:test-cover-incompressibility}}
\end{figure}

\begin{itemize}
\item  
It adds sets of vertices $R$ and $B$ to $G$,
where $R$ contains a vertex corresponding 
to each vertex in $R'$ and
$B$ contains two vertices corresponding
to each vertex in $B'$.
Formally, 
$R = \{r_i |\ i \in [|R'|]\}$
and
$B = \{b^{\circ}_i, b^{\star}_i |\ i \in [|B'|]\}$.

\item The reduction adds a bit representation gadget to locate set $B$.
It adds vertices such that for any pair $b^{\circ}_j, b^{\star}_j$, the supplementary vertices adjacent to them are identical.
\begin{itemize}
\item The reduction sets $p := \lceil \log(|B|/2) \rceil+1$ and
for every $i \in [p]$, it adds two vertices $z_{i, 1}$ and $z_{i, 2}$ and edge $(z_{i, 1},  z_{i, 2})$.
It adds all $z_{i,1}$'s to $R$ and all $z_{i, 2}$'s to $B$.
\item For every integer $j \in [|B|/2]$, let $\bit(j)$ denote the binary representation of $j$ using $p$ bits.
Connect $b^{\circ}_{j}, b^{\star}_j \in B$ with $z_{i, 1}$ if the $i^{th}$ bit (going from left to right) in $\bit(j)$ is $1$.
\item It add two vertices $z_{0, 1}$ and $z_{0, 2}$, and edge $(z_{0, 1}, z_{0, 2})$.
It adds $z_{0,1}$ to $R$ and $z_{0, 2}$ to $B$.
It also makes every vertex in $B$ adjacent with $z_{0, 1}$.
Let $\bitrep(B)$ be the collection of vertices of type
$z_{i,1}$ for $j \in \{0\}\cup [p]$.
\end{itemize}

\item The reduction adds edges 
across $R$ and $B$ as follows:
For every pair of vertices $r'_i \in R'$
and $b'_j \in B'$, 
\begin{itemize}
\item if $r'_i$ is \emph{not} adjacent with $b'_j$ then  it adds both edges
$(r_i, b^{\circ}_j)$ and $(r_i, b^{\star}_j)$ to $E(G)$,
and
\item if $r'_i$ is adjacent with $b'_j$
then it adds $(r_i, b^{\circ}_j)$ only, i.e.,
it does not add an edge with endpoints $r_i$ and 
$b^{\star}_j$.
\end{itemize}

\item Finally, it adds an isolated blue vertex $b_0$.
\end{itemize}
This completes the construction of 
$G$.
The reduction sets 
$$k = k' + |\bitrep(B)| =  k' + \log(|B|/2) + 1 = \calO(|B|),$$
and returns $(G, \langle R, B \rangle, k)$ as an instance of 
\TCPB.

It is easy to verify that $(G, \langle R, B \rangle, k)$ is a valid instance
of \TCPB.
We present a brief overview of the proof of correctness.
Suppose $(G', \langle R', B' \rangle, k')$ is a \yes-instance
of \textsc{Red-Blue Dominating Set} and let set $S' \subseteq R'$
be a solution.
We argue that $S' \cup \bitrep(B)$
is a locating-dominating set of $G$ of size at most $k$.
Note that red vertex $z_{0,1}$ ensures
that all the pendant vertices that are adjacent
with some vertex in $\bitrep(B)$ has unique 
neighbour in $S$. 
Hence, apart from the vertex pairs of the form
$(b^{\circ}_j, b^{\star}_j)$, every other pair of vertices 
is resolved by vertices in  $\bitrep(B)$.
Now, by the construction of $G$,
a vertex $r_i \in R$ resolves a pair vertices $b_j^{\circ}$
and $b_j^{\star}$
if and only if $r'_i$ and $b'_j$ are adjacent with each other. 
This concludes the forward direction of the proof.
In the reverse direction, note that $b_0$ is an isolated 
blue vertex.
Hence, all the red vertices in $\bitrep(B)$
are forced into any solution, say $S_1$, as they are unique red vertices
that are adjacent with some blue vertex.
Hence, by the property mentioned above, it is easy to see that 
$S_1 \setminus \bitrep(B)$ is a desired dominating
set in graph $G'$.
This concludes the overview.

The proof of Thereom~\ref{thm:loc-dom-set-incompressibility}
for \TCPB follows with identical arguments as presented in 
Section~\ref{sec:incompressibility}.
Recall that \textsc{Red-Blue Dominating Set},
parameterized by the number of blue vertices, does not admit
a polynomial kernel unless $\NP \subseteq \coNP/poly$.
See, for example, \cite[Lemma~$15.19$]{cygan2015parameterized}.
Moreover, when parameterized by 
the number of red vertices, it does not admit
a polynomial kernel unless $\NP \subseteq \coNP/poly$.
See, for example, \cite[Exercise~$15.4.10$]{cygan2015parameterized}.
This, along with the arguments that are standard to parameter
preserving reductions, imply that
\TCPB, parameterized by 
$(i)$ the number of items $|U|$ and the solution size $k$, or
$(ii)$ the number of sets $|\calF|$
does not admit a polynomial kernel,
unless $\NP \subseteq \coNP/poly$.

%% file: conclusion.tex
\section{Conclusion}
\label{sec:conclusion}

We investigated the structural parameterization 
of \LD and \TCPB.
We presented several results about the algorithmic complexity of 
these two problems.
However, our main technical contribution is an \FPT\ algorithm
when parameterized by the vertex cover number, 
a linear kernel when parameterized by the feedback edge set number,
and a non-compressibility result.
Our first two results imply that the double-exponential 
lower bound when parameterized by treewidth in~\cite{DBLP:journals/corr/abs-2402-08346}
cannot be extended to larger parameters like vertex cover number
and feedback edge set number.
The reduction used to prove the third result
also provides simplified proofs of some known results.

We do not know whether our $2^{\calO({\vc}\log {\vc})}n^{\calO(1)}$-time  \FPT\ algorithm
for \LD\ parameterized by vertex cover (and similarly, the algorithm
for \TCPB parameterized by $|U|$) is optimal, or whether there exist single-exponential algorithms.
Another open question in this direction, is whether this algorithm can be generalized to the parameter distance to cluster.
As our linear kernel for parameter feedback edge set number is not explicit, it would be nice to obtain a concrete kernel.